\documentclass[aps,prb,superscriptaddress,floatfix,preprint]{revtex4-1}
\usepackage{amssymb}
\usepackage{amsmath}
\usepackage{graphicx}
\usepackage{color}
\usepackage{siunitx}

\begin{document}

\title{\Large $4\pi$-periodic Josephson supercurrent in HgTe-based topological Josephson junctions}

\author{J.~Wiedenmann}
\altaffiliation{All three authors contributed equally to this work, email: erwann.bocquillon@physik.uni-wuerzburg.de}

\author{E.~Bocquillon}
\thanks{All three authors contributed equally to this work, email: erwann.bocquillon@physik.uni-wuerzburg.de}
\affiliation{Physikalisches Institut (EP3), Universit\"at W\"urzburg, Am Hubland, D-97074 W\"urzburg, Germany}

\author{R.S.~Deacon}
\thanks{All three authors contributed equally to this work, email: erwann.bocquillon@physik.uni-wuerzburg.de}
\affiliation{Advanced Device Laboratory, RIKEN, 2-1 Hirosawa, Wako-shi, Saitama, 351-0198, Japan}
\affiliation{Center for Emergent Matter Science, RIKEN, 2-1 Hirosawa, Wako-shi, Saitama, 351-0198, Japan}

\author{S.~Hartinger}
\affiliation{Physikalisches Institut (EP3), Universit\"at W\"urzburg, Am Hubland, D-97074 W\"urzburg, Germany}

\author{O.~Herrmann}
\affiliation{Physikalisches Institut (EP3), Universit\"at W\"urzburg, Am Hubland, D-97074 W\"urzburg, Germany}

\author{T.M.~Klapwijk}
\affiliation{Kavli Institute of Nanoscience, Faculty of Applied Sciences, Delft University of Technology, Lorentzweg 1, 2628 CJ Delft, The Netherlands}
\affiliation{Laboratory for Quantum Limited Devices, Physics Department, Moscow State Pedagogical University, 29 Malaya Pirogovskaya St. Moscow 119992, Russia}

\author{L.~Maier}
\affiliation{Physikalisches Institut (EP3), Universit\"at W\"urzburg, Am Hubland, D-97074 W\"urzburg, Germany}

\author{C.~Ames}
\affiliation{Physikalisches Institut (EP3), Universit\"at W\"urzburg, Am Hubland, D-97074 W\"urzburg, Germany}

\author{C.~Br\"une}
\affiliation{Physikalisches Institut (EP3), Universit\"at W\"urzburg, Am Hubland, D-97074 W\"urzburg, Germany}

\author{C.~Gould}
\affiliation{Physikalisches Institut (EP3), Universit\"at W\"urzburg, Am Hubland, D-97074 W\"urzburg, Germany}

\author{A.~Oiwa}
\affiliation{The Institute of Scientific and Industrial Research, Osaka University 8-1 Mihogaoka, Ibaraki, Osaka 567-0047, Japan}

\author{K.~Ishibashi}
\affiliation{Advanced Device Laboratory, RIKEN, 2-1 Hirosawa, Wako-shi, Saitama, 351-0198, Japan}
\affiliation{Center for Emergent Matter Science, RIKEN, 2-1 Hirosawa, Wako-shi, Saitama, 351-0198, Japan}

\author{S.~Tarucha}
\affiliation{Center for Emergent Matter Science, RIKEN, 2-1 Hirosawa, Wako-shi, Saitama, 351-0198, Japan}
\affiliation{Department of Applied Physics, University of Tokyo, 7-3-1 Hongo, Bunkyo-ku, Tokyo, 113-8656, Japan}

\author{H.~Buhmann}
\affiliation{Physikalisches Institut (EP3), Universit\"at W\"urzburg, Am Hubland, D-97074 W\"urzburg, Germany}

\author{L.W.~Molenkamp}
\affiliation{Physikalisches Institut (EP3), Universit\"at W\"urzburg, Am Hubland, D-97074 W\"urzburg, Germany}

\maketitle

\noindent

{\bf The Josephson effect describes the generic appearance of a supercurrent in a weak link between two superconductors. Its exact physical nature however deeply influences the properties of the supercurrent. Detailed studies of Josephson junctions can reveal microscopic properties of the superconducting pairing (spin-triplet correlations, $d$-wave symmetry) or of the electronic transport (quantum dot, ballistic channels). In recent years, considerable efforts have focused on the coupling of superconductors to topological insulators, in which transport is mediated by topologically protected Dirac surface states with helical spin polarization (while the bulk remains insulating). Here, the proximity of a superconductor is predicted to give rise to unconventional induced $p$-wave superconductivity, with a doublet of topologically protected gapless Andreev bound states, whose energies varies $4\pi$-periodically with the superconducting phase difference across the junction. In this article, we report the observation of an anomalous response to rf irradiation in a Josephson junction with a weak link of the 3D topological insulator HgTe. The response is understood as due to a $4\pi$-periodic contribution to the supercurrent, and its amplitude is compatible with the expected contribution of a gapless Andreev doublet.}

\begin{figure}[h!]
\centerline{\includegraphics[width=1\textwidth]{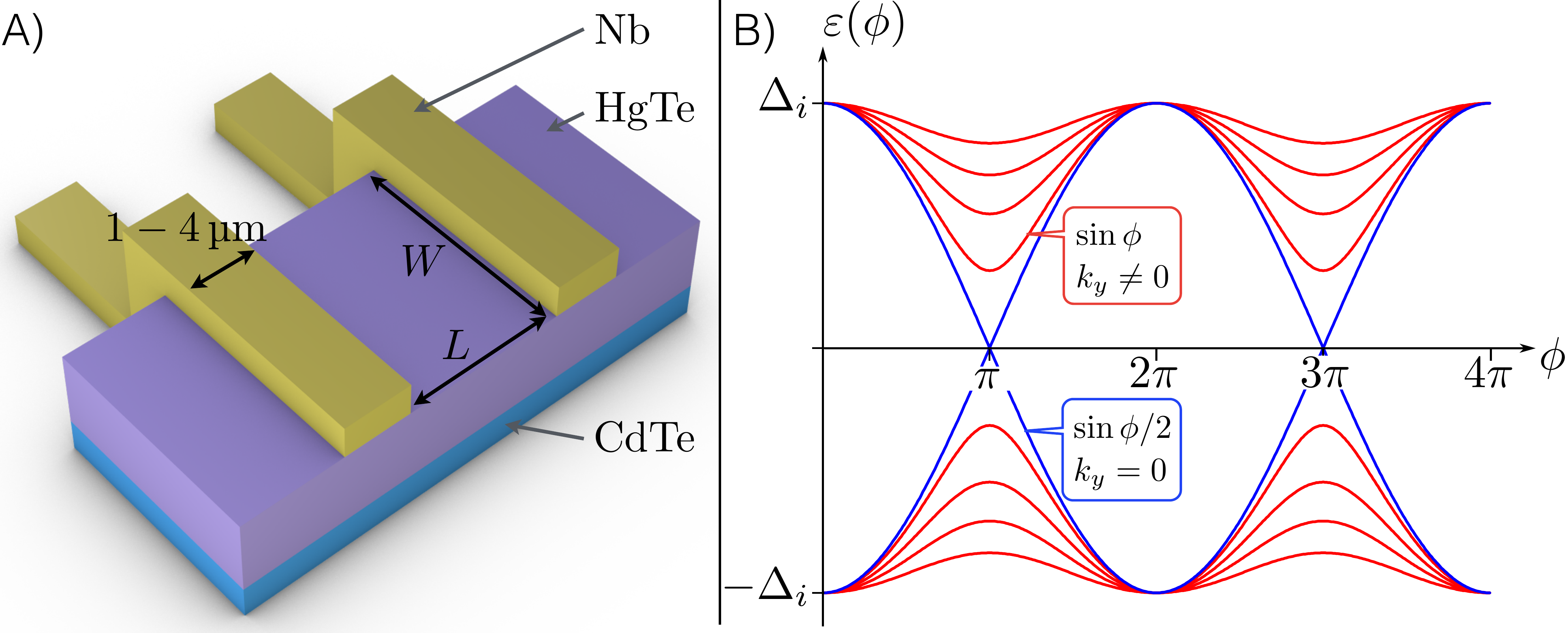}}
\caption{{\bf Geometry of the Josephson junction and predicted Andreev spectrum -} (A) Artist view of the Josephson junction. Mesa stripes of HgTe (represented in mauve) are patterned on the CdTe substrate (blue), with a width $W=\SI{2}{\micro\meter}$. Nb contacts (in yellow) are added at the surface, with a width of of 1 to \SI{4}{\micro\meter}, separated by a variable distance $L$. (B) Typical energy spectra $\varepsilon(\phi)$ of the Andreev bound states in a 3D TI based junction, as a function of the phase difference $\phi$ in the JJ. In blue is depicted the gapless $4\pi$-periodic topological mode, corresponding to transverse momentum $k_y=0$ and contributing to the $4\pi$-supercurrent $I_{4\pi}\sin{\phi/2}$. In red, gapped modes correspond to $k_y\neq0$ and contribute to the $2\pi$-supercurrent $I_{2\pi}$ (containing $\sin\phi$ and higher order harmonics).} \label{Fig:Andreev}
\end{figure}

The helical nature of the topological surface states, where the spin is locked perpendicular to the momentum \cite{Fu2007}, is predicted to give rise to exotic superconductivity when coupled to the conventional pairing potential of a $s$-type superconductor. The broken spin rotation symmetry allows the appearance of triplet $p$-wave correlations and of gapless Andreev bound states, regardless of the microscopic details of the theoretical model\cite{Fu2008,Olund2012,Beenakker2013,Tkachov2013}. In a 3D topological insulator (TI) based Josephson junction (JJ), in which superconductivity is induced by the proximity effect of a $s$-wave superconductor (see Fig.\ref{Fig:Andreev}A), Andreev bound states appear in the induced gap $\Delta_i$ that can be pictured as (see Fig.\ref{Fig:Andreev}B) a single topological Andreev doublet (depicted in blue) that occurs at transverse momentum $k_y=0$ and is immune to back-scattering (thus has perfect transmission), and non-topological oblique modes ($k_y\neq0$, depicted in red) that are expected to have lower transmissions. The topological protection of the zero mode constitutes a superconducting analogue to Klein tunneling\cite{Tkachov2013}. As depicted in Fig.\ref{Fig:Andreev}B, the peculiarity of this topological doublet is its $4\pi$-periodicity (or equivalently a contribution $I_{4\pi}\sin\phi/2$ to the supercurrent) with respect to the superconducting phase difference $\phi$ across the junction \cite{Kitaev2001, Beenakker2013, Zhang2014}. The non-ambiguous observation of such gapless states is regarded as an important experimental signature of the unconventional superconductivity in topological insulators, but no robust evidence has been reported yet \cite{Veldhorst2012, Oostinga2013, Maier2015, Kurter2014,Galletti2014,Finck2014}. A major hindrance could be the coexistence of residual bulk conductance or of a large number of gapped conventional modes \cite{Finck2014,Galletti2014}. Furthermore the finite lifetime of the positive energy branch prevents observation of a $4\pi$-periodic Josephson effect in stationary measurements as various relaxation mechanisms can restore a $2\pi$-periodicity for the current-phase relation (CPR) \cite{Kwon2004, SanJose2012,Pikulin2012, Badiane2013}. The $4\pi$-periodic Josephson effect can therefore be unveiled more easily by the dynamics of the junction. To reveal the periodicity of the Josephson supercurrent, an rf driving current $I_{rf}$ is added to the dc drive to induce the so-called Shapiro steps\cite{Shapiro1963}. When the dynamics of a conventional JJ is phase-locked to the rf drive, steps of constant voltage appear in the $I$-$V$ characteristic of the junction for voltages $V_n=nhf/2e$ where $n\in\mathbb{Z}$ is the step index. However, for a purely $4\pi$-periodic supercurrent, only a sequence of even steps should be observed. In the case of nanowires \cite{Rokhinson2012}, signs of the disappearance of the first step ($n=1$) have been reported and attributed to a theoretically expected topological phase transition driven by a magnetic field along the axis of the nanowire, although the topological state in the nanowires has yet to be identified in the normal transport regime. Even though it is known that other systems (JJ between $p$/$d$-wave superconductors \cite{Tanaka1997,Kwon2004}) could lead to similar effects without the need for topological protection, such effects have yet to be observed. In this article, we study HgTe, a genuine 3D topological insulator whose topological properties have been established independently\cite{Bruene2011, Bruene2014}, and observe an anomalous doubled Shapiro step appearing at low frequency (equivalently a missing $n=1$ step). While several other mechanisms (non-linearities, capacitance effects, higher harmonics in the current phase relation  \cite{Renne1974,Valizadeh2008}) are known to cause the {\it appearance} of {\it additional} subharmonic steps in the Shapiro response, to our knowledge, only the existence of a $4\pi$-periodic contribution $I_{4\pi}\sin\phi/2$ in the total supercurrent can be responsible for the {\it disappearance} of odd steps \cite{Dominguez2012}.

\begin{figure}[h!]
\centerline{\includegraphics[width=1\textwidth]{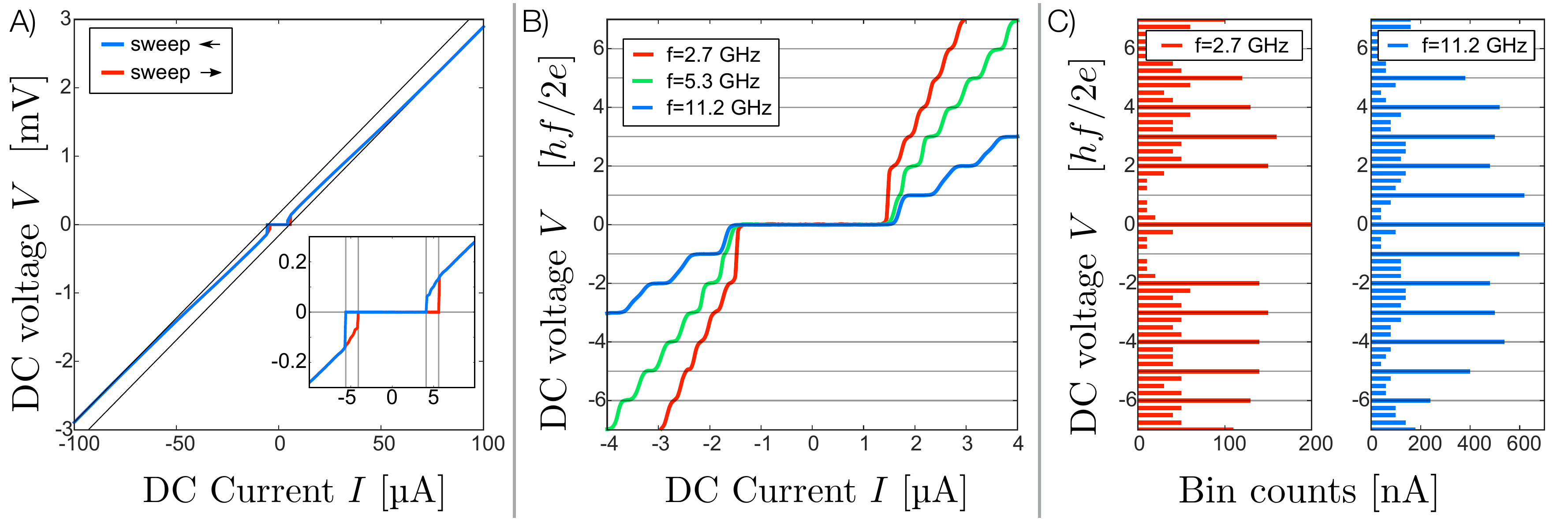}}
\caption{{\bf $I$-$V$ curves of the $L=\SI{150}{\nano\meter}$ junction) -} (A) $I$-$V$ curve in the absence of rf irradiation for the two sweep directions, taken at base temperature $T\simeq \SI{30}{\milli\kelvin}$. The asymptotes (grey solid lines) do not cross the origin, emphasizing the presence of an excess current. (Inset) Detailed view of the $I$-$V$ curve, that exhibits hysteresis between the upward and downward sweep direction. (B) Shapiro steps for three different frequencies measured at $T\simeq \SI{800}{\milli\kelvin}$. The plotted voltage scale is in normalized units $hf/2e$ to highlight the formation of Shapiro steps in the $I$-$V$ curve in the presence of rf irradiation. For a high frequency $f=\SI{11.2}{\giga\hertz}$ (blue line), all steps are clearly visible for voltages $V_n=n\frac{hf}{2e},\, n\in\mathbb{Z}$ (up to $|n|>12$, but only the first three are shown for the sake of clarity). For an intermediate frequency ($f=\SI{5.3}{\giga\hertz}$, blue line), the first step ($n=1$) is noticeably reduced. At low frequency ($f=\SI{2.7}{\giga\hertz}$, red line), the first step is fully suppressed, while all other steps remain visible. (C) Bar plots obtained by binning the measurement data according to voltage, for $f=\SI{2.7}{\giga\hertz}$ and \SI{11.2}{\giga\hertz}. The Shapiro steps appear as peaks in the bin counts for $V_n=n\frac{hf}{2e},\, n\in\mathbb{Z}$. While all steps are visible for $f=\SI{11.2}{\giga\hertz}$ (right graph), the first Shapiro step ($n=1$) is absent at $f=\SI{2.7}{\giga\hertz}$ (left graph). } \label{Fig:DCmeas}
\end{figure}

Our devices are fabricated from coherently strained undoped HgTe layers of \num{65} to \SI{90}{\nano\meter} thickness, epitaxially grown on a CdTe substrate. The band inversion of HgTe enforces the existence of topological surface states, while strain opens a gap ($\simeq \SI{22}{\milli\electronvolt}$) in the bulk of the material \cite{Fu2007a}. Previous work has highlighted the high quality of the topological states in this material\cite{Bruene2011, Oostinga2013, Bruene2014}. Quantized Hall plateaus are routinely observed, which demonstrate that transport occurs exclusively through the surface states, without any detectable parallel conductance from the bulk. The mobility and charge density, relevant for our experiments, are evaluated from a Hall-bar produced separately from the same wafer as the junctions, and yield typically $\mu= 1-\SI{3e4}{\centi\meter\squared\per\volt\per\second}$, and $n_e= 3-\SI{7e11}{\per\centi\meter\squared}$. From these values, we extract a mean free path of $l\simeq \SI{200}{\nano\meter}$. The JJs are fabricated by depositing niobium contacts at the surface of a HgTe mesa, using standard sputtering and lift-off techniques (see Supplementary information). The geometry is shown in Fig.\ref{Fig:Andreev}A. Each superconducting contact has a width of 1 to \SI{4}{\micro\meter}, the HgTe weak link has a width of $W=\SI{2}{\micro\meter}$ (corresponding to the width of the mesa stripe) and a variable length $L$ ranging from \SI{150}{\nano\meter} to \SI{600}{\nano\meter}. From the electron density $n_e$, we evaluate the number of transport modes $N=\frac{Wk_F}{\pi}\simeq100$. The niobium of the contacts has a critical temperature of $T_c\simeq 8$ K, slightly lower than that of bulk Nb (9.2 K). A typical $I$-$V$ curve obtained at \SI{30}{mK} is presented in Fig.\ref{Fig:DCmeas}A) and exhibits hysteresis, as commonly reported \cite{Courtois2008,Oostinga2013}. We find that the critical current of devices with the same dimensions varies by about 30\%, which underlines the reproducibility and quality of the fabrication process. A recurring feature in all devices is the presence of an excess current in the $I$-$V$ curve (see Fig.\ref{Fig:DCmeas}A). For high biases, the $I$-$V$ curves become linear with an asymptote which does not go through the origin but is shifted towards higher currents. This excess current is understood as due to the fact that electrons in an energy window near the superconducting gap carry twice as much current due to Andreev reflections \cite{Blonder1982,Klapwijk1982}. It thus illustrates the presence of Andreev reflections at both S-TI interfaces. Such an excess current has been previously observed for superconducting point-contacts \cite{Scheer2001,Weitz1978}, but is not commonly reported in thin film structures presumably due to the presence of elastic scattering. This emphasizes the high quality and reproducibility of our devices in agreement with our previous observations \cite{Oostinga2013,Maier2015,Sochnikov2014}.

We now turn to the study of the response of these devices to rf irradiation and highlight the existence of a $4\pi$-periodic supercurrent. To this end, we focus on three devices produced from the same wafer, for which the width of the junction is set to $W=\SI{2}{\micro\meter}$, for nominal lengths of $L=\,$\SIlist{150;400;600}{\nano\meter} (see Supplementary information). The experiment described below has been repeated on more than ten devices, made out of three different wafers with similar characteristics, in three different measurement setups, all yielded similar results (see Supplementary information).
 
In order to observe the Shapiro steps, the sample is irradiated with a radio-frequency excitation via a coaxial line, the open end of which is adjusted to be around 1 mm from the sample. In this geometry, frequencies in the range of 2 to 12 GHz are easily accessible, but the rf power supplied to the sample is not calibrated. Under rf irradiation, we observe the appearance of Shapiro steps in the $I$-$V$ characteristic at quantized voltages $V_n=nhf/2e$, where $n\in\mathbb{Z}$ is the step index\cite{Shapiro1963}. In contrast to the standard Josephson junction response, with steps at each $n$, we find at lower frequency that the $n=1$ step is missing. To illustrate this anomalous Shapiro response of our junctions, we present three $I$-$V$ curves corresponding to three different excitation frequencies in Fig.\ref{Fig:DCmeas}B (for the junction with $L=\SI{150}{\nano\meter}$). The applied rf power is chosen such that all curves display similar critical currents, the full range of rf power will be discussed later. For a high frequency $f=\SI{11.2}{\giga\hertz}$, one typical $I$-$V$ curve is plotted as a blue line (with voltage normalized to $hf/2e$). Several steps are clearly visible with step height $hf/2e$. At lower frequencies $f=\SI{5.3}{\giga\hertz}$ (green line), higher order steps are visible but a clear reduction of the amplitude of the $n=1$ step occurs. For a frequency of $f=\SI{2.7}{\giga\hertz}$ (red line), this first odd step is fully suppressed, showing an anomalous first step at $hf/e$. The presence or absence of the $n=1$ can be conveniently detected by binning the measurement data according to the voltage (with a $0.25\frac{hf}{2e}$ bin size). The resulting histograms of the voltage $V$ are presented as bar plots in Fig.\ref{Fig:DCmeas}C. For $V_n=nhf/2e$ with $n$ integer, Shapiro steps appear as peaks in the bin counts, the amplitude of which then reflects the length of the current step (in nA). For $f=\SI{11.2}{\giga\hertz}$ (right graph), all steps emerge clearly from the background. On the left side, for $f=\SI{2.7}{\giga\hertz}$, the peak at $V=hf/2e$ is absent, reflecting the suppression of the $n=1$ Shapiro step. This anomalous behaviour of the Shapiro steps constitutes the main finding of this article. Below, we carefully analyze its origin and conclude that it indicates the existence of a $4\pi$-periodic contribution to the supercurrent. 

\begin{figure}[h!]
\centerline{\includegraphics[width=0.8\textwidth]{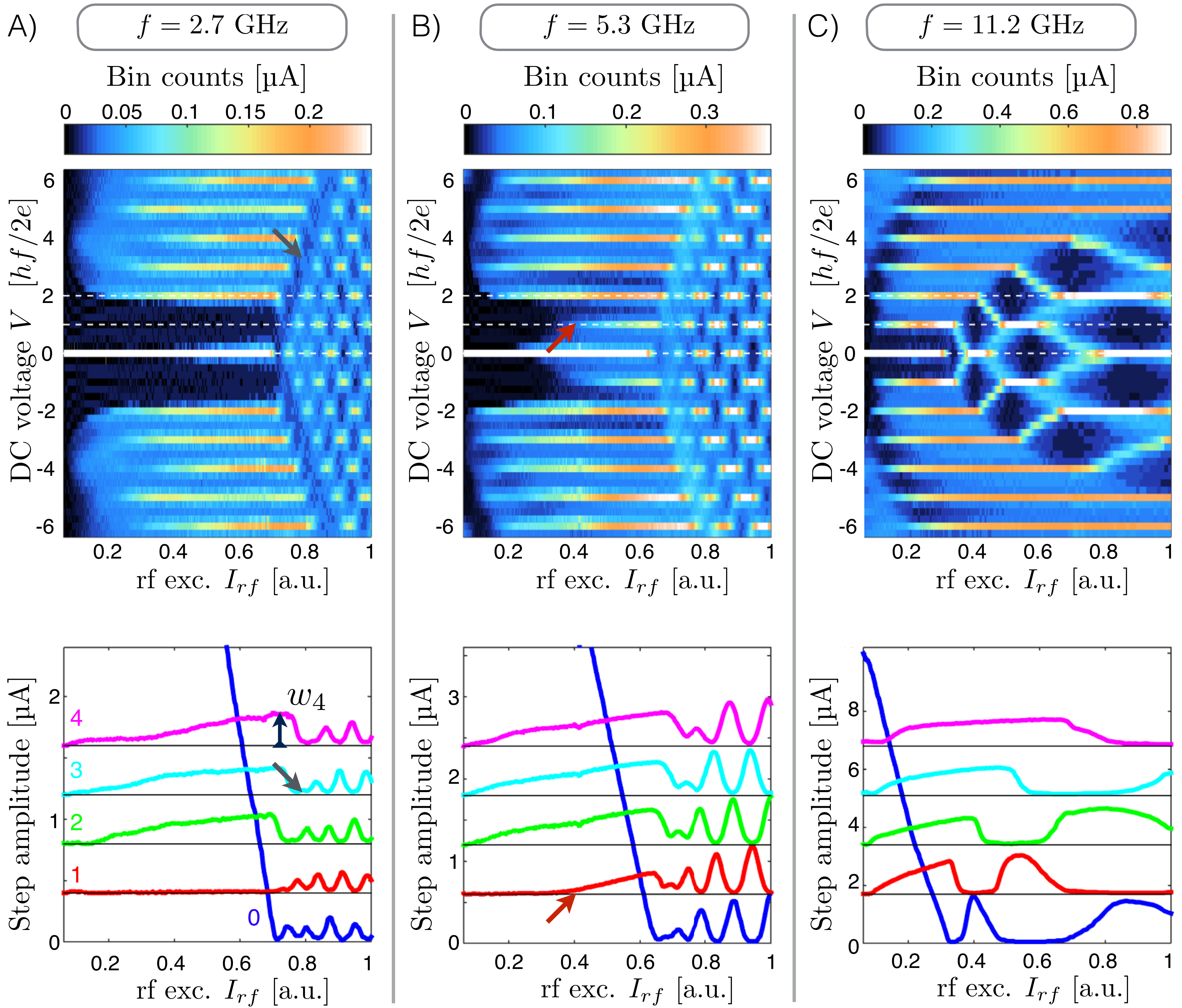}}
\caption{{\bf 2D plots of the bin counts and Shapiro step amplitudes for the $L=\SI{150}{\nano\meter}$ junction) -} Differential conductance is plotted as a function of the voltage $V$ (normalized by the step size at frequency $f$  $hf/2e$) and rf current $I_{rf}$ for rf excitation frequencies $f=\SI{2.7}{\giga\hertz}, \SI{5.3}{\giga\hertz}, \SI{11.2}{\giga\hertz}$ in (A), (B) and (C) respectively. (upper panels) Shapiro steps are identified as maxima for constant voltages $V_n$ (white dashed lines emphasize $n=0,1,2$). For $f=\SI{11.2}{\giga\hertz}$, all steps are visible. When frequency is lowered ($f=\SI{5.3}{\giga\hertz}$), the first odd step ($n=1$) is absent up to a rf excitation indicated by the red arrow. Finally, at $f=\SI{2.7}{\giga\hertz}$, the first step is completely invisible up to the crossing point which marks the beginning of the oscillatory regime at high rf currents. A dark fringe (indicated by a dark grey arrow) is observed at finite voltages in the oscillating pattern concomitant with the missing $n=1$ step. (lower panels) Horizontal line-cuts through the previous colormaps give access to the amplitudes of steps 0 to 4. While all Shapiro steps are clearly visible at high frequencies, the step $n=1$ progressively disappears as $f$ decreases. From these plots, we access the maximum widths $w_n$ of each step (see the example of $w_4$ at $f=\SI{2.7}{\giga\hertz}$). For clarity, the different curves are offset by 0.4, 0.6 and \SI{1.7}{\micro\ampere} for (A), (B), (C) respectively.} \label{Fig:2DPlots}
\end{figure}

We now examine the crossover from high to low frequency, for which the first odd Shapiro step $n=1$ progressively disappears. To this end, we scan the presence of Shapiro steps for a range of rf powers at fixed frequencies and generate two-dimensional color plots of the bin counts at the voltage $V$ (which indicates the current-height of the Shapiro step when present) as a function of the voltage $V$ and rf current $I_{rf}$. As shown in Fig.\ref{Fig:2DPlots} (for the junction with $L=\SI{150}{\nano\meter}$), such plots reveal the presence of Shapiro steps as maxima at constant quantized voltages (horizontal lines). Let us first examine measurements taken at $f=\SI{11.2}{\giga\hertz}$. (Fig.\ref{Fig:2DPlots}C). At $I_{rf}=0$, a single maximum at $V=0$ reflects the presence of a supercurrent. As $I_{rf}$ increases, Shapiro steps progressively appear, starting from low values of $n$, while the amplitude of the supercurrent ($n=0$) decreases and eventually vanishes. At higher powers, the steps show an oscillatory pattern, reminiscent of Bessel functions occurring in the voltage bias case \cite{Tinkham2004,Russer1972}. Horizontal linecuts at constant voltages give access to the amplitude of the first steps ($n=0,1,2,3,4$), presented in the lower panels of Fig.\ref{Fig:2DPlots} as a function of rf current $I_{rf}$. For high frequencies such as $f=\SI{11.2}{\giga\hertz}$, our device exhibits the conventional behavior which is seen in various other systems (carbon nanotubes \cite{Cleuziou2007}, graphene\cite{Heersche2007}, or $\rm Bi_2Se_3$ \cite{Galletti2014} weak links), that always (regardless of frequency) show a clear presence of the $n=1$ step. The case of atomic contacts (with a few ballistic highly-transparent modes) is particularly well understood, and also exhibits a strong $n=1$ Shapiro resonance in excellent agreement with theoretical models \cite{Chauvin2006,Cuevas2002}. In the Supplementary Information, we provide additional measurements on graphene-based devices (another example of 2D Dirac material), that also shows this standard behavior in all accessible regimes.

In contrast to the conventional Shapiro features commonly reported, our HgTe-based junctions exhibit a very clear vanishing of the first step $n=1$ when the excitation frequency $f$ is decreased. Measurements at $f=\SI{5.3}{\giga\hertz}$ show that the first step is suppressed below a certain value of $I_{rf}$ (indicated by the red arrow), and that it is completely absent at $f=\SI{2.7}{\giga\hertz}$. In the oscillatory regime at higher rf currents, a suppressed first oscillation (dark fringe indicated by the dark grey arrow) becomes clearly visible at low frequency, demonstrating the range of influence of the vanishing first step on the rest of the pattern. In the lower panels, a complete suppression of the first step or disturbances in the oscillations at higher rf currents can similarly be observed. This crossover has been observed on all working devices, up to 800 mK, which is the highest stable temperature accessible in our fridge. In some cases, hysteretic behavior at low temperatures hinders the observation of low-index steps (see Supplementary Information). However, importantly, biasing instabilities and sudden current switches (such as the ones observed in the hysteretic regime) can be excluded as a mechanism for the missing $n=1$ step, as the same features are seen in measurements of a junction in which bistability is suppressed by a shunt resistor (see Supplementary Information).

In opposition to a missing $n=1$ step, additional subharmonic steps (for $n=p/q$ fractional value) are often observed \cite{Dubos2001,Chauvin2006} as a consequence of non-linearities, capacitance effects or higher harmonics in the CPR. Such higher harmonics have been predicted\cite{Tkachov2013} and detected\cite{Sochnikov2014} in our junctions. At higher frequencies, we indeed observe half-integer steps ($n=1/2, 3/2,...$, see Supplementary Information) but they clearly appear in a different regime from where we observe the missing $n=1$ step.

\begin{figure}[h!]
\centerline{\includegraphics[width=0.7\textwidth]{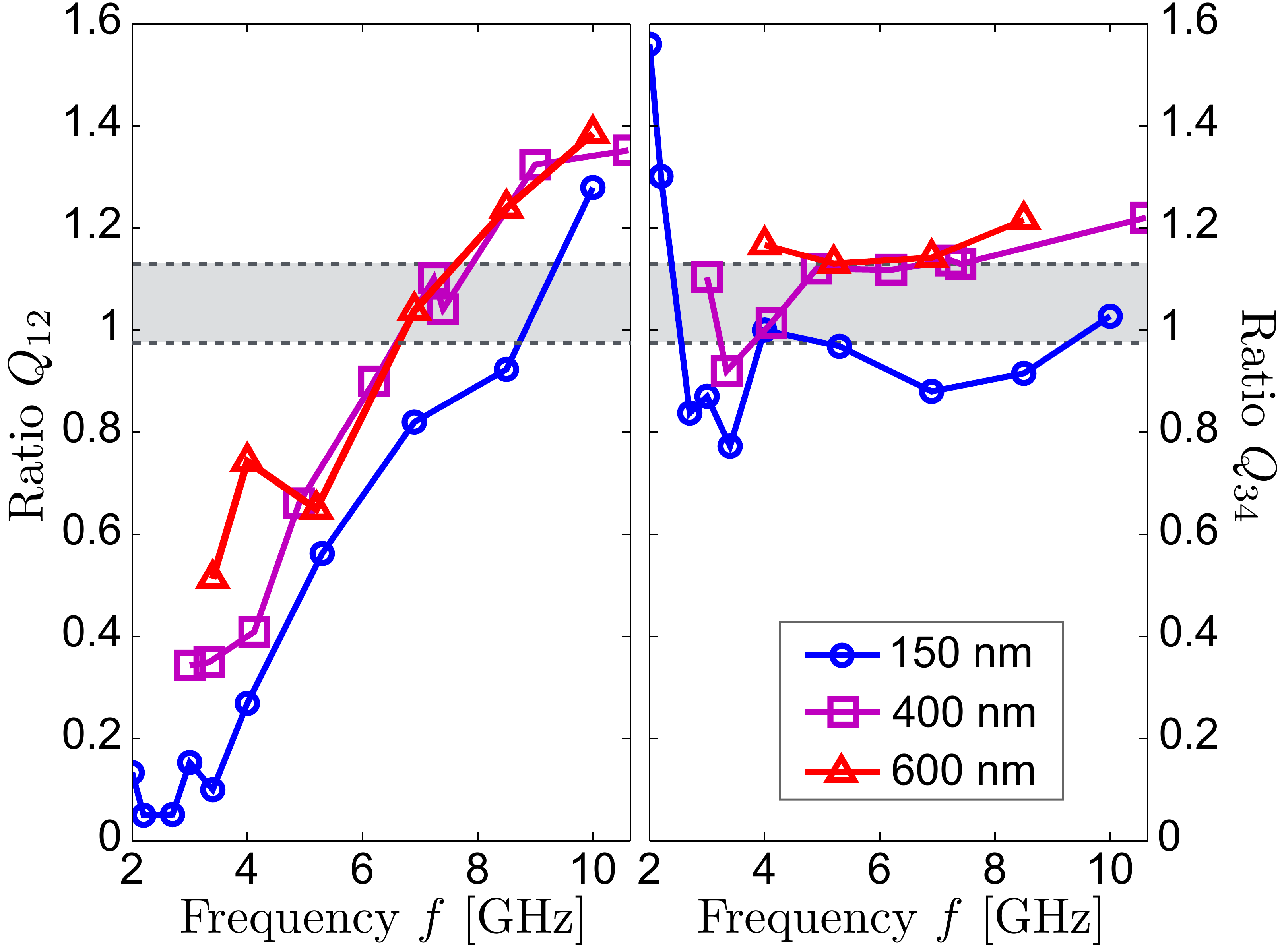}}
\caption{{\bf Ratios of step widths $Q_{12}$ and $Q_{34}$ vs frequency $f$ -} For each length $L$ of the JJ, we calculate the ratios of step amplitudes $Q_{12}=w_1/w_2$ and $Q_{34}=w_3/w_4$ and plot them as a function of the rf frequency. $Q_{12}$ shows a very clear decrease as frequency $f$ is lowered. A minimum around 0.05 is obtained for the 150 nm junction, but we observe that this minimum tends to increase with the length $L$ of the junction. In contrast, even if the measurements show some scattering, the ratio of higher order steps $Q_{34}$ does not show significant variation. For comparisons, we evaluated $Q_{12}$ and $Q_{34}$ from two conventional RSJ models, and show the results as a grey area (see Supplementary Information).} \label{Fig:Ratios}
\end{figure}

The presence of a $4\pi$-contribution in the supercurrent $I_{4\pi}\sin\phi/2$ is the only known mechanism to result in the observed doubling of the Shapiro step size. As already mentioned, microscopic models based on Bogoliubov-de Gennes equations have predicted such a $4\pi$-periodic contribution in the CPR \cite{Fu2008,Olund2012,Beenakker2013,Tkachov2013}, which originates from the presence of a gapless topological Andreev doublet. This anomalous CPR can then be supplemented with the Josephson equation on the time-evolution of the phase difference to simulate the dynamics of such a system. This dynamics is captured in the extended RSJ model of Dominguez {\it et al.} \cite{Dominguez2012}. It takes into account the presence of a $\sin\phi/2$ contribution in the supercurrent and explains the crossover between the two frequency regimes by the highly non-linear dynamics of the junction. When a small $4\pi$-periodic contribution $I_{4\pi}\sin\phi/2$ is superposed on a large $2\pi$-periodic supercurrent $I_{2\pi}\sin\phi$ in the current-phase relation, the latter dominates the high-frequency Shapiro response, but the weak $4\pi$ contribution is revealed at low frequencies by doubled Shapiro steps (see Supplementary Information). Doubled Shapiro steps are observed only when the driving frequency $f$ becomes smaller than the characteristic frequency $f_{4\pi}=\frac{2eR_n I_{4\pi}}{h}$ (with $R_n$ the normal state resistance of the device). This frequency scale based on the amplitude of the $4\pi$ supercurrent is expected to be much smaller than the typical Josephson frequency scale $f_J=\frac{2eR_n I_c}{h}$ ($f_J\simeq\SI{53}{\giga\hertz}$ for the 150 nm long junction), as $I_{4\pi}\ll I_{2\pi}\simeq I_c$. In order to estimate $I_{4\pi}$, we introduce two indicators $Q_{12}$ and $Q_{34}$ as follows. From the maximum amplitude of the first lobe of each step, denoted by $w_n, n\in \mathbb{Z}$, (see Fig.\ref{Fig:2DPlots}A where the measurement is indicated for the $n=4$ step), we define and compute the ratios $Q_{12}=w_1/w_2$, $Q_{34}=w_3/w_4$, and plot them as a function of the rf excitation frequency (Fig.\ref{Fig:Ratios}). Despite some scattering, we observe a clear decrease of $Q_{12}$ towards 0 with decreasing frequency, while $Q_{34}$ remains constant around 1, for all lengths. For the shortest junction (150 nm) $Q_{12}$ reaches a value of 0.05 around 2 GHz, and the first step $n=1$ is invisible. For comparison, we have also plotted the boundaries (grey dashed lines) between which the ratios $Q_{12}$ and $Q_{34}$ vary in the standard RSJ model\cite{McCumber1968,Russer1972} (with only a $\sin\phi$ component in the supercurrent, see Supplementary Information). While the ratio $Q_{34}$ remains close to grey region, the behavior of $Q_{12}$ is not properly described. Assuming the validity of the above criterion, one can evaluate the number of $4\pi$-periodic channels. We estimate $f_{4\pi}=4.5-\SI{5}{\giga\hertz}$ and $I_{4\pi}=250-\SI{300}{\nano\ampere}$ for the 150 nm junction, and $f_{4\pi}=\SI{4}{\giga\hertz}$ and $I_{4\pi}=50-\SI{70}{\nano\ampere}$ for the longer junctions (400 and 600 nm). One can compare these values with the maximum supercurrent carried by one channel \cite{Beenakker1991}, given by $e\Delta_i/\hbar$ per channel where $\Delta_i$ can be estimated from the decay of $I_c$ with temperature (see Supplementary information). With $\Delta_i=\SI{0.35}{\milli\electronvolt}$ (150 nm) and $\Delta_i=0.1-\SI{0.15}{\milli\electronvolt}$ (400 and 600 nm), we estimate that the $4\pi$-contribution amounts to that of 1-3 channels which is compatible with the presence of one topological mode in our system, despite uncertainties on the exact value of $f_{4\pi}$ and $\Delta_i$. Results on three other junctions with different parameters have been compiled (see Supplementary Information) and are consistent with this estimate.

Finally, one might also suspect that the $4\pi$-periodic contribution stems from Landau-Zener transitions occurring at the anticrossing (for $\phi=\pi\,[2\pi]$), causing some highly-transparent $2\pi$-periodic states to behave effectively as $4\pi$-periodic, in the absence of truly $4\pi$-periodic modes. In a single mode model\cite{Dominguez2012}, one can numerically show that the quantization of the Shapiro steps is lost when the Landau-Zener tunneling probability is lower than 1: the Shapiro steps split in two and depart from their quantized values $V_{n}$, ($n$ even), and eventually disappear for probabilities below 0.7. We do not experimentally observe such effects in any accessible regime. Assuming the validity of this specific model, an upper bound on the possible energy splitting $2\delta$ between positive and negative energy branches can be evaluated from the Landau-Zener transition probability. Given our experimental resolution, we obtain the upper bound $\delta\leq\SI{9}{\micro\electronvolt}$ for the 400 and 600 nm junctions, and $\delta\leq\SI{18}{\micro\electronvolt}$ for the 150 nm one (see Supplementary Information). This is much smaller than the energy scale given by the temperature ($\SI{70}{\micro\electronvolt}$ at 800 mK) and corresponds in both cases to a transmission $\geq 0.994$. Besides, there are no reports of missing odd Shapiro steps due to Landau-Zener transitions in highly ballistic junctions reported to date.

Interestingly, only the first step $n=1$ is missing\footnote{This is similar to that reported for etched InSb nanowire devices \cite{Rokhinson2012}}, and not the following odd steps $n=3,5,...$. This feature is not well understood, but one possible explanation is given by enhanced relaxation (due for example to coupling to the continuum of states above the superconducting gap), with a characteristic time scale that decreases as voltage increases \cite{SanJose2012,Houzet2013}. However, most models assume a voltage bias of the junction and a more detailed analysis of the current bias case is needed.

To conclude, we have presented robust evidence for a $4\pi$-periodic contribution to the supercurrent flowing in Josephson junctions based on the 3D topological insulator HgTe. The consistency of the measurements in Hall bars and JJs signals that our devices are well-controlled, with well-defined proximity-induced superconducting HgTe contacts connected via a ballistic HgTe surface. Under rf irradiation, a suppression of the first Shapiro step is observed at low frequencies and low magnetic fields, for a wide range of temperatures (up to 800 mK), which we attribute to the existence of a $4\pi$-periodic component in the supercurrent. The study of its order of magnitude and of Landau-Zener transitions reveal that these experimental observations are compatible with the presence of a few $4\pi$-periodic gapless Andreev bound states. Such states would likely stem from the topologically protected gapless Andreev bound states, but could also originate from trivial ballistic states. Further investigations are required to conclusively demonstrate the relationship of these observations to Majorana physics \cite{Alicea2012, Beenakker2013}. Besides, these observations with a 3D TI of strained HgTe are very encouraging for future experiments in which the weak link would consist of narrow HgTe quantum wells that exhibit the quantum spin Hall effect \cite{Konig2007}, in which the total number of transport modes should be reduced to a few. 

{\bf Acknowledgments:}

We gratefully acknowledge G. F\`eve and B. Pla\c cais for the loan of amplifiers, as well as M. Houzet, J. Meyer, H. Pothier, F. Hassler, A. Brinkman, C.W.J. Beenakker, A. Akhmerov, P. Burset, E.M. Hankiewicz, G. Tkachov and B. Trauzettel for enlightening discussions. This work is supported by German Research Foundation (DFG-JST joint research project ´Topological Electronics´ and the Leibniz Program), the EU ERC-AG program (Project 3-TOP), the Elitenetzwerk Bayern program “Topologische Isolatoren” and the DARPA MESO project. R.S.D. acknowledges support from Grants-in-Aid for Young Scientists B (No. 26790008). T.M.K. is financially supported by the European Research Council Advanced grant No.339306 (METIQUM) and by the Ministry of Education and Science of the Russian Federation under Contract No.14.B25.31.007. S.T. acknowledges financial support from Grants-in-Aid for Scientific Research S (No. 26220710) and JST Strategic International Cooperative Program. E.B., T.M.K. and L.W.M. gratefully thank the Alexander von Humboldt foundation for its support.

\clearpage

\clearpage

%make figure S1 etc...
\renewcommand{\thefigure}{S\arabic{figure}}
%reset counter
\setcounter{figure}{0}

\clearpage
\begin{center}

\Large{$4\pi$-periodic Josephson supercurrent in HgTe-based topological Josephson junctions\\
--\\
Supplementary Material}
\normalsize
\end{center}

{\bf 
In this supplementary online text, we first detail the sample preparation and give further results on the properties of the Josephson junctions. Then additional results are presented, obtained from different HgTe-based devices. In particular, the issue of subharmonic steps and hysteresis are discussed. In a third part, we present the Shapiro response of graphene-based devices measured for comparison with the HgTe samples. Finally, we discuss elements of theory on RSJ simulations (with and without $4\pi$-periodic supercurrent) and Landau-Zener transitions.
}

\section{Additional characterization measurements}

\subsection{Sample preparation and layer characterization}

Bulk HgTe layers are grown by molecular beam epitaxy on a CdTe substrate. Prior to the fabrication of the JJs, the transport properties of each layer are characterized by the measurement of longitudinal and transverse (Hall) resistance in a Hall bar geometry. From the longitudinal resistance at zero magnetic field, one can extract the mobility of the layer, while the density is obtained from a linear fit of the Hall resistance between 0 and 500 mT. The layers being very similar to the ones presented in references \onlinecite{Bruene2011, Bruene2014}, we refer the interested reader to these references where the measurements are discussed in detail.\\
Using a Ti/SiO$_2$ etch mask, the samples are patterned via Ar$^+$ ion beam milling to obtain \SI{2}{\micro\meter} wide HgTe stripes. The Nb contacts (60 nm Nb + Al/Ru cap) are sputtered on top of the HgTe layer, after a short Ar$^+$-milling step in order to remove any adsorbant or reaction product left on the HgTe mesa after exposure to air. The top-layer of HgTe below the contact may be disordered due to the deposition of the Nb contacts. We assume that this top-surface is proximitized by interaction with the Nb, but that the HgTe weak link between the contacts remains in the ballistic regime as highlighted in the main text.

\begin{figure}[h!]
\centering\includegraphics[width=1\textwidth]{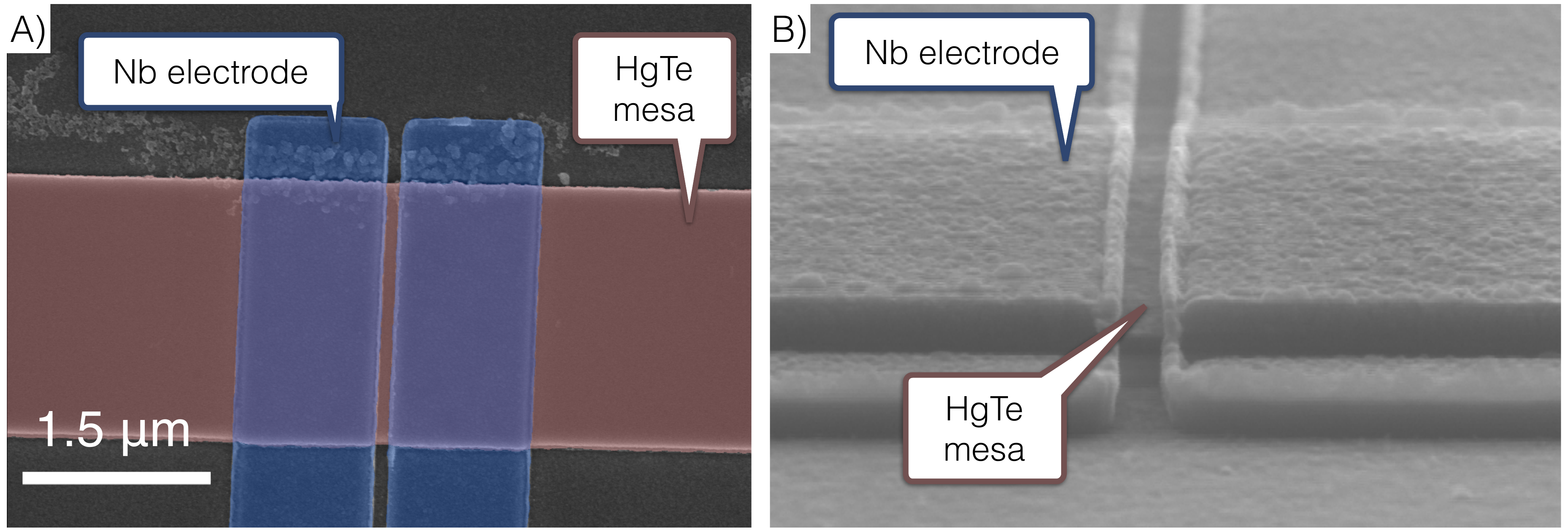}
\caption{{\bf SEM pictures of two typical devices -} (A) Colorized SEM pictures, highlighting the HgTe mesa (in red) and the Nb electrodes (in blue), in the lateral junction geometry presented in Fig.1 (main text). (B) Close-up on the junction itself. Nb sidewalls are visible on the electrodes, and are a consequence of sputter deposition on the sidewalls of the resist pattern used to define contacts.} \label{Fig:SOMImageJJ}
\end{figure}

In Fig.\ref{Fig:SOMImageJJ}, SEM pictures of a typical junction are shown. From these SEM pictures, it is observed that the nominal length (used in the article) probably overestimates the physical length of the JJs. We evaluate the physical length to be 60 to 70 nm smaller than the nominal length.

\subsection{Temperature dependence}

\begin{figure}[h!]
\centering\includegraphics[width=1\textwidth]{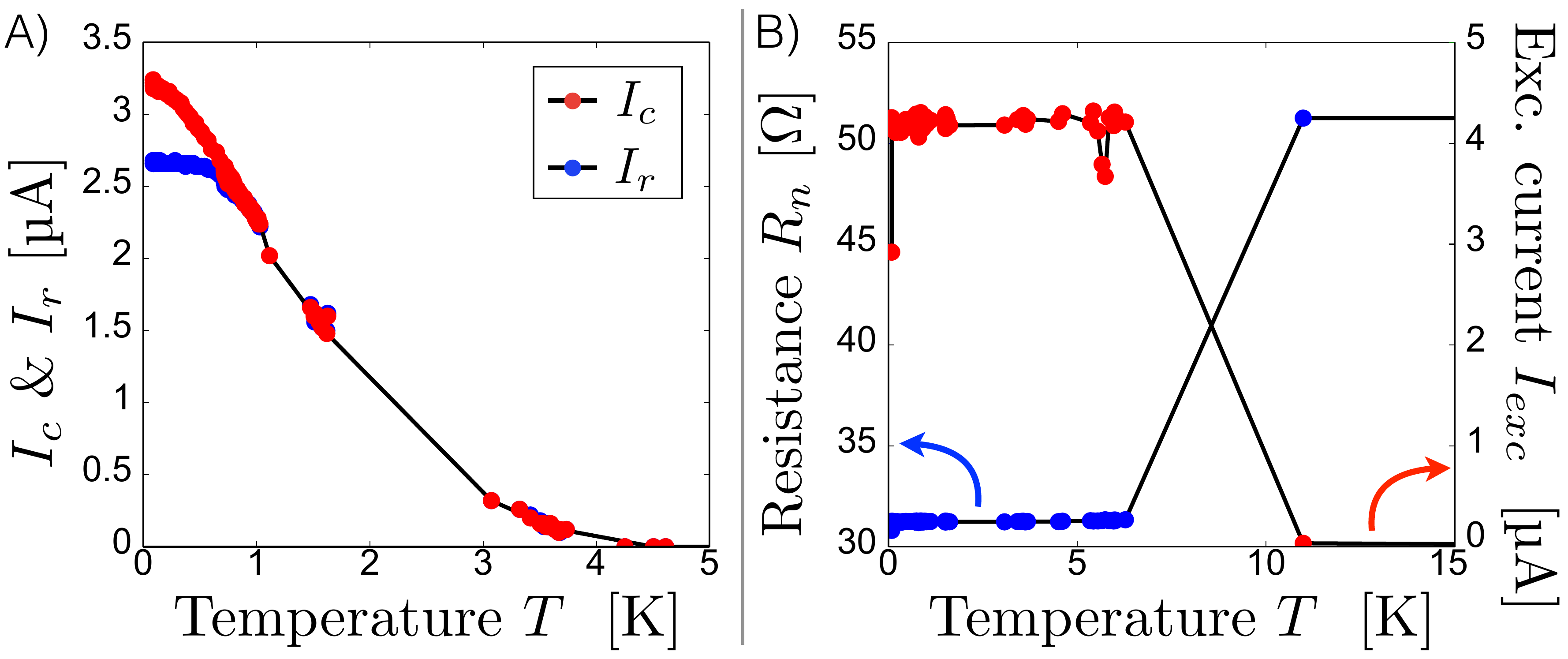}
\caption{{\bf Temperature dependence -} (A) Temperature dependence of the critical and retrapping current $I_c$ and $I_r$. The hysteretic region is seen for $T<\SI{800}{\milli\kelvin}$ typically, for which $I_r<I_c$. (B) Temperature dependence of the excess current $I_{exc}$ and normal state resistance $R_n$. The superconducting phase transition in the Nb contacts is seen here as a shift in the resistance $R_n$, and typically takes place at $T_c\simeq\SI{8}{\kelvin}$. Simultaneously, the excess current $I_{exc}$ exhibits a jump from zero to a finite value, thus reflecting the presence of Andreev reflections at the interfaces of the junction for temperatures $T<T_c$.} \label{Fig:SOMTempDep}
\end{figure}

We present in Fig.\ref{Fig:SOMTempDep} the temperature dependence of four quantities. In Fig.\ref{Fig:SOMTempDep}A, the critical current $I_c$ and retrapping current $I_r$ are shown (in red and blue dots respectively). Both decay on a typical scale of 1-2 \si{\kelvin}. The hysteretic region, characterized by $I_c>I_r$ takes place at roughly $T\leq\SI{800}{\milli\kelvin}$ in most samples. Beyond this temperature, $I_c=I_r$ and no hysteresis is observed in the $I-V$ curve. In Fig.\ref{Fig:SOMTempDep}B, the excess current $I_{exc}$ and the normal state resistance $R_n$ of the junction are plotted as a function of temperature $T$. First, the resistance $R_n$ (red dots) exhibits a jump (from \SI{31}{\ohm} to \SI{51}{\ohm}) indicating the superconducting transition of the bulk Nb contacts. These measurements were taken in a dilution fridge for which stable temperature control was only possible below \SI{1}{\kelvin}. Higher temperature measurements were collected while the system warmed up allowing only a few reliable measurement points such that the exact transition temperature is not known. From other measurements during cool down, we obtain a critical temperature around $T_c\simeq\SI{8}{\kelvin}$, slightly smaller than that of pure high-quality Nb (\SI{9.2}{\kelvin}) and compatible with the measurements presented here. Simultaneously, the excess current is measured and exhibits a jump from zero to a finite value (around \SI{4}{\micro\ampere}) as the Nb contacts become superconducting. This signals the presence of Andreev reflections at both interfaces of the junctions for all temperatures $T<T_c$.

\begin{figure}[h!]
\centering\includegraphics[width=\textwidth]{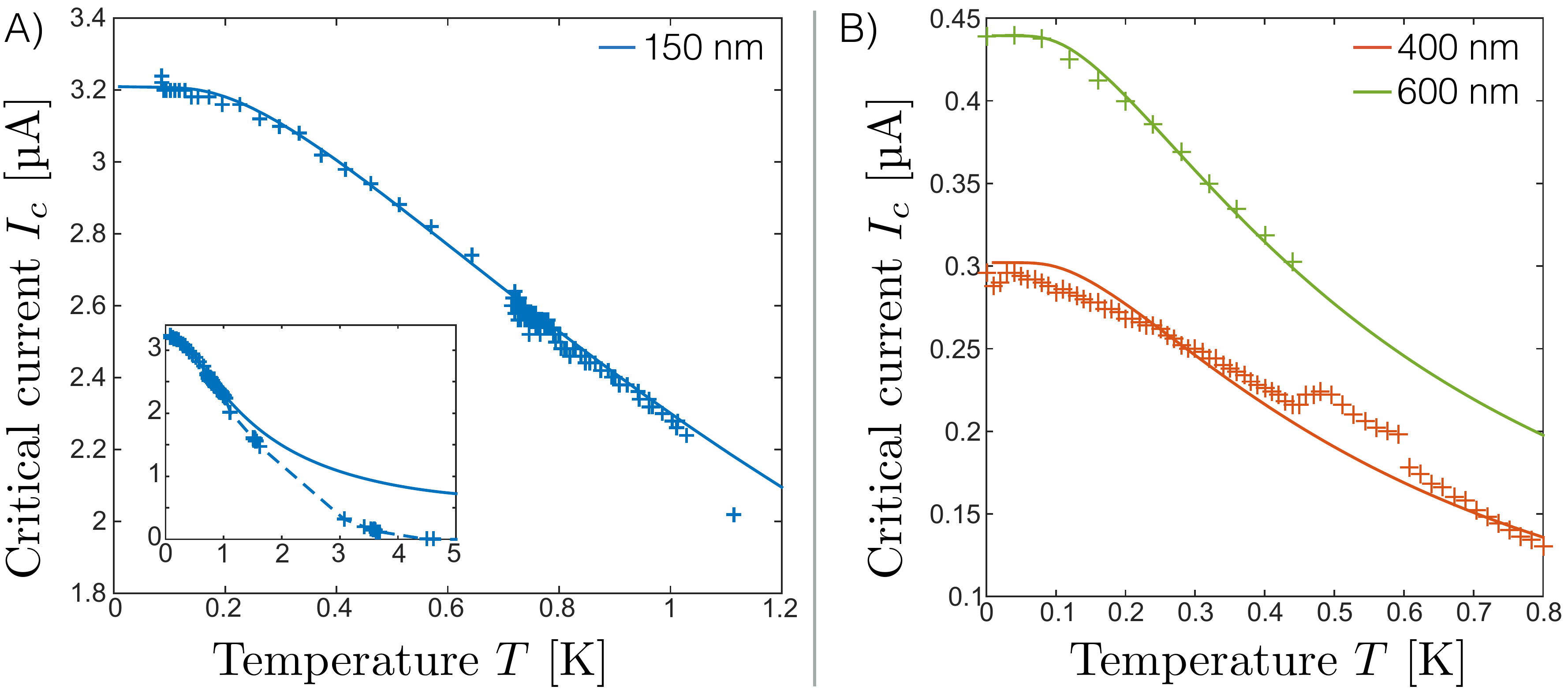}
\caption{{\bf Fit of of the critical current vs. temperature -} (A) Experimental data for the critical current $I_c(T)$ as a function of temperature $T$ is presented as blue + symbols, for the 150 nm JJ. The solid line is obtained using Eq.(\ref{Eq:TkachovFit}) for three-dimensional topological Josephson junctions, with $\Delta_i=\SI{0.35}{\milli\electronvolt},\, N=117,\, Z=0.1$. (inset) Complete temperature range. Beyond 1K, data is taken as the fridge warms up, resulting in inaccurate measurements of $T$. The fit departs from experimental data as the perturbative model breaks down. (B) The temperature dependence of $I_c(T)$ is presented for the 400 and 600 nm junctions, in red and green respectively. Experimental data is showed as + symbols while solid lines present fits with parameters $\Delta_i=\SI{0.13}{\milli\electronvolt},\, N=48,\, Z=1$ (600 nm) and $\Delta_i=\SI{0.13}{\milli\electronvolt},\, N=33,\, Z=1$ (400 nm). } \label{Fig:SOMTempDepFits}
\end{figure}

The temperature dependence of the junction with $L=\SI{150}{\nano\meter}$ presented in Fig.\ref{Fig:SOMTempDep} can be analyzed using the approach developed by Tkachov {\it et al.}\cite{Tkachov2013}. Each mode (indexed by the transverse component of the wavevector $k_y$) is described by an angle-dependent transmission $T_\theta$ where $\sin\theta=k_y/k_F$. The distribution of transmission $T_\theta$ reflects Klein tunneling through a barrier (characterized by a parameter $Z$ in a BTK-like approach\cite{Blonder1982}), with topological protection of the mode $k_y=0, (T_{\theta=0}=1)$.
This model also takes into account induced superconductivity in the HgTe reservoir using a McMillan tunneling approach\cite{McMillan1968}, which allows for the evaluation of the induced gap $\Delta_i$. Unfortunately, this perturbative approach breaks down for high temperature when the induced gap $\Delta_i$ becomes comparable to the Nb gap $\Delta_{Nb}$ (that decreases with temperature).
\begin{eqnarray}
I_c(T)&=&\frac{e\Gamma}{4\hbar}\sum_{k_y} \sin\phi\cos\theta\frac{T_\theta(1-\gamma+\frac{5}{2}\gamma^2-\frac{3}{2}\gamma^2T_\theta\sin^2\frac{\phi}{2})}{1-T_\theta\sin^2\frac{\phi}{2}}\nonumber\\
&&\quad\tanh\Gamma\frac{(1-\gamma+\gamma^2)(1-T_\theta\sin^2\frac{\phi}{2})^{1/2}+\frac{1}{2}\gamma^2(1-T_\theta\sin^2\frac{\phi}{2})^{3/2}}{2k_BT}\label{Eq:TkachovFit}\\
T_\theta&=&\frac{\cos^2\theta}{1-\sin^2\theta/(1+Z^2)},\, \gamma=\Gamma/\Delta_{Nb},\,\Delta_i=\Gamma(1-\gamma+\frac{3}{2}\gamma^2)\nonumber
\end{eqnarray}

Eq.(\ref{Eq:TkachovFit}) has three fit parameters: the tunneling strength $\Gamma$ (or equivalently the induced gap $\Delta_i$), the barrier parameter $Z$ and the number of modes $N$. As seen on Fig.\ref{Fig:SOMTempDepFits}A, the agreement below 1 K is very good for the $L=\SI{150}{nm}$ junction, but the fit diverges rapidly at high temperature. We obtain the following fitting parameters $\Gamma=\SI{0.4}{\milli\electronvolt}, \Delta_i=\SI{0.35}{\milli\electronvolt},\, Z=0.1, N=117$. The number of modes is then in agreement with independent estimates from Hall bar measurements, and yields an estimate of $\Delta_i$ that can be used to estimate the number of $4\pi$-periodic modes (see section \ref{Summary}).

For longer junctions ($L=400 - \SI{600}{\nano\meter}$), the agreement is not as good as for the 150 nm long junction presented here. The evaluation of the temperature scale over which the decay of $I_c$ is observed gives an estimate of $\Delta_i=0.1-\SI{0.15}{\milli\electronvolt}$ lower than for the junction with $L=\SI{150}{nm}$. The amplitude of the critical current $I_c$ decreases rapidly with length, and yields a number of modes $N<50$ which seems unreasonably small given the estimate of the density from a separate Hall-bar, or from the value of $N$ in the 150 nm long device on the same sample. It could indicate that these junctions depart from the short junction limit $l\ll\xi$ where $\xi$ is the coherence length in the system.
The natural coherence length $\xi_0=\frac{\hbar v_F}{\pi\Delta_i}$ is typically between 250 and 1000 nm in our system\cite{Sochnikov2014}. For systems with mean free path $l\sim\xi_0$, the relevant length is in fact $\xi=\sqrt{\xi_0 l}$ in the range of 250 to 550 nm which is compatible with our findings.

\subsection{Excess current}

The excess current in our data is indeed strikingly present in all our samples. We take it as an additional indication of the systematic reproducibility of our data within a large variety of samples of different fabrication runs and measured in different cryostats at different locations. The excess current in a Josephson junction is obtained for voltages beyond 2$\Delta$, as discussed by Blonder {\it et al.}\cite{Blonder1982}, and reaches in principle twice the value for a NS contact. However, the amplitude of the excess current depends on elastic scattering at the interfaces and as such is not a sufficient measure. The presence of gapless modes can in principle be detected by a halved onset on the excess current: the asymptotic regime is reached for a bias eV on the order of the gap $\Delta$ and not 2$\Delta$ in the conventional case. 

\begin{figure}[h!]
\centering\includegraphics[width=0.9\textwidth]{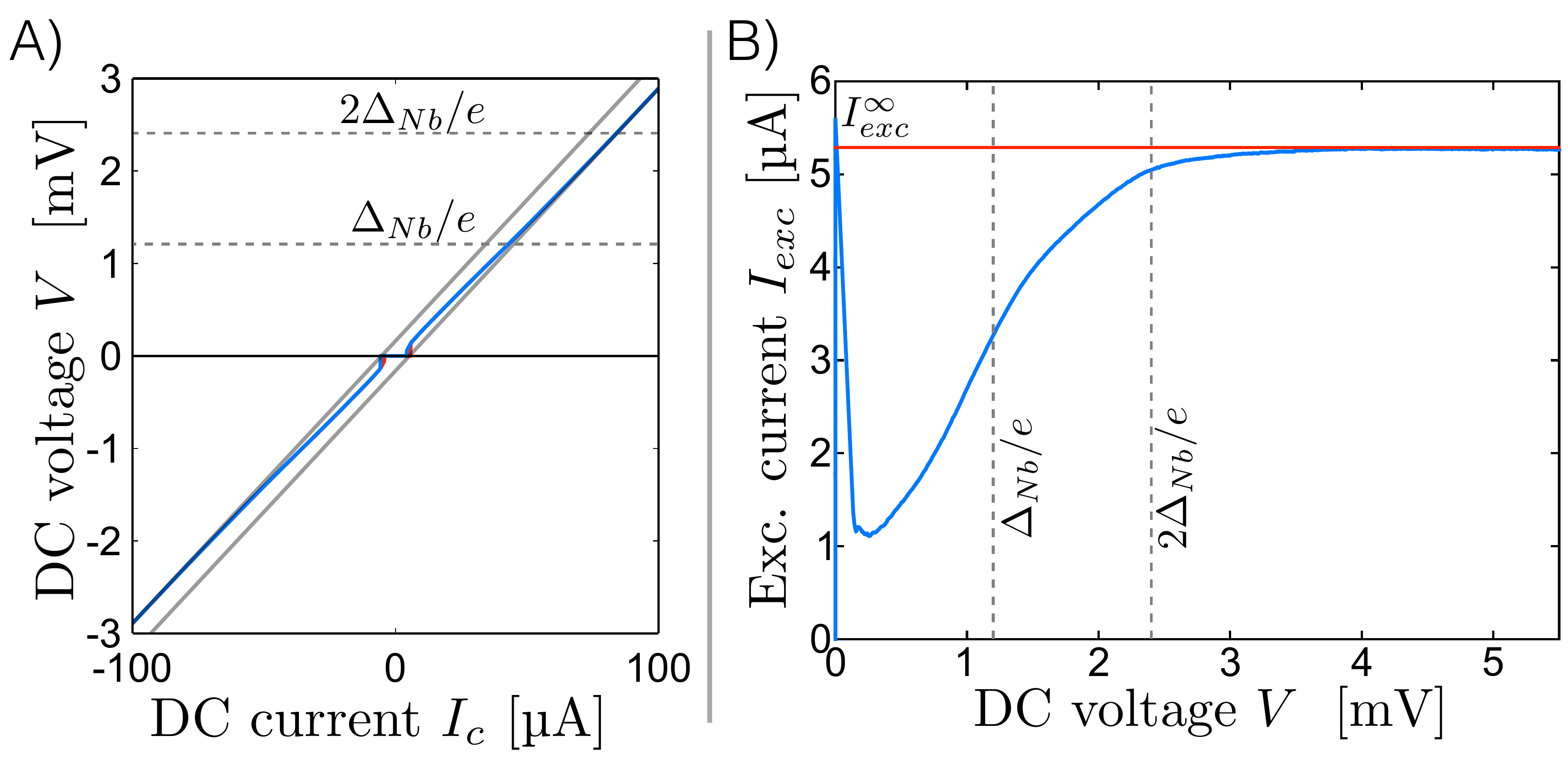}
\caption{{\bf Excess current and midgap states -} A) The $I$-$V$ curve of the 150 nm junction is presented in blue and red line (for the two sweep directions). Two asymptotes are presented as a grey line, that do not cross the origin, demonstrating the presence of an excess current. Two horizontal dashed line have been added as guidelines and represent possible onsets at $eV\simeq\Delta_{Nb}$ and $eV\simeq2\Delta_{Nb}$. B) Excess current $I_{exc}(V)=I-V/R_n$ as a function of voltage $V$. Inflexions are visible around $V\simeq\SI{1.4}{\milli\volt}$ and $V\simeq\SI{2.4}{\milli\volt}$. At high bias, $I_{exc}(V)\to I_{exc}^\infty=\SI{5.3}{\micro\ampere}$.} \label{Fig:SOMExcessCurrent}
\end{figure}

To analyze more precisely this behavior, we plot in the right panel of Fig.\ref{Fig:SOMExcessCurrent} the difference $I_{exc}(V)=I-V/R_n$, with $I_{exc}(V)\to I_{exc}^\infty$ for $eV\gg2\Delta_{Nb}$. We observe two inflexions around $V\simeq\SI{1.4}{\milli\volt}$ and $V\simeq\SI{2.4}{\milli\volt}$, that could be related to $\Delta_{Nb}$ and $2\Delta_{Nb}$. However, these features are relatively weak, and no clear transition at $eV\simeq\Delta_{Nb}$ is observed.

The fact that we cannot observe a clear asymptotic regime at $\Delta$ is however not unexpected as the previous model is only directly applicable to N-S contacts, in which a single superconducting contact is involved and thus the dynamics of Josephson effect is not present. In a Josephson S-S contact, the observed currents and voltages below $2\Delta$ are mixed with the time-averaged quantities of the Josephson effect. The $I$-$V$ curve thus reveals a time-averaged part due to the Josephson effect and a static part due to the Andreev reflection processes. To get experimental access to the « excess current » part, it is necessary to separate these contributions. The contributions due to the Josephson effect are strongest at $V=0$ and at finite voltages close to $V=0$ but gradually decay in amplitude with increasing voltage. In contrast the excess current is present out to high voltages beyond $2\Delta$ and perhaps beyond $\Delta$. Careful experimental work is therefore needed to disentangle these two processes from the experimental data.
Furthermore, the theory has been developed for nanowires (single-band and short junction limit) with a simple representation of induced superconductivity, by a unique gap $\Delta$. The fact that our system is a 2D surface state rather than a nanowire is expected to have little consequence. However, we expect to have a very different density of states due to the presence of two superconducting gaps in proximity with each other (the bulk Nb gap $\Delta_{Nb}$ and the induced gap $\Delta_i$). The literature\cite{Fagas2005,Kopnin2014} shows for example the presence of features at both $\Delta_{Nb}$ and $\Delta_i$, as well as different densities of states in regions of energy $\Delta_i<\epsilon<\Delta_{Nb}$ and $\epsilon>\Delta_{Nb}$. The onset of the excess current in our measurements is of the same order of magnitude as $\Delta_{Nb}\simeq \SI{1}{\milli\electronvolt}$, and the role of the induced gap (evaluated around $\Delta_i\simeq 0.1-\SI{0.4}{\milli\electronvolt}$) remains experimentally not accessible.

\section{Additional results on $\rm HgTe$-based junctions}

\subsection{Frequency dependence and half-integer steps}
\label{Subharm}

\begin{figure}[h!]
\centering\includegraphics[width=0.9\textwidth]{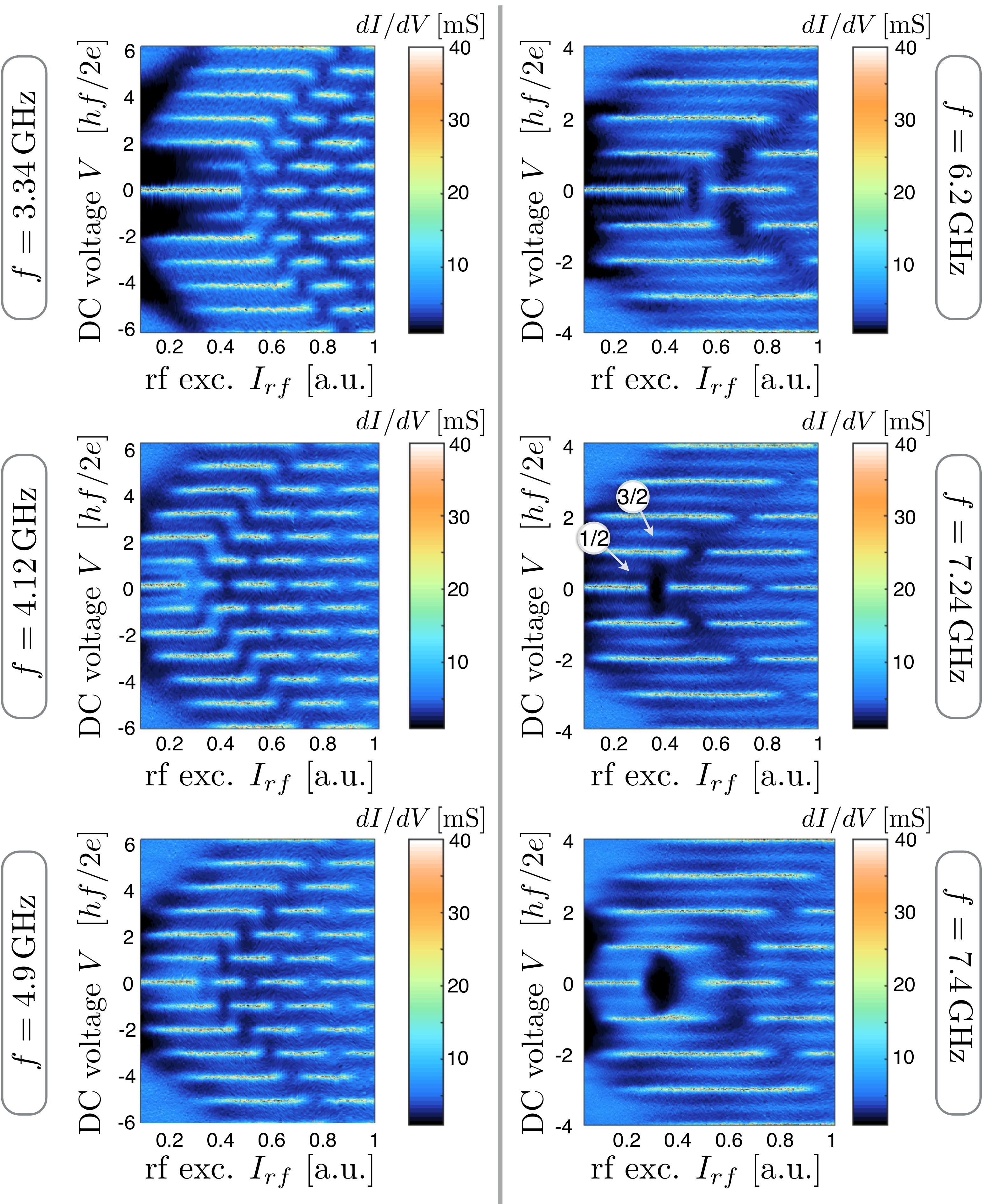}
\caption{{\bf Frequency dependence of the Shapiro response -} The differential conductance $dI/dV$ is plotted as a colorscale, as a function of the dc voltage $V$ and rf current drive $I_{rf}$ for different frequencies. The progressive appearance of the initially (partially) missing $n=1$ step is observed. At high frequencies ($f\geq\SI{6.2}{\giga\hertz}$), subharmonic steps $n=1/2,3/2,5/2...$ become visible (see section \ref{Subharm}).} \label{Fig:SOMShapiroRock400nm}
\end{figure}

In Fig.\ref{Fig:SOMShapiroRock400nm}, we present datasets measured for a different sample than the one presented in the main text with length $L=\SI{400}{\nano\meter}$. In particular, we show the transition from a doubled Shapiro step (at $f=\SI{3.34}{\giga\hertz}$) to a regime in which the first step is fully recovered. At high frequencies (for $f\geq\SI{6.2}{\giga\hertz}$), we observe the appearance of new half-integer steps, at voltages given by the step indexes $n=1/2,3/2, 5/2,...$. 
To visulaize more clearly the subharmonic steps, we introduce a different way of visualizing our data. The differential conductance $dI/dV$ is plotted as a colorscale, as a function of the dc voltage $V$ and rf current drive $I_{rf}$. Thus, Shapiro steps appear as maxima ($dI/dV$ diverges) for constant voltages, similarly to what is seen in the bin counts presented in the main text. In such plots, the information on the step amplitude is lost, but subharmonic steps of small amplitudes become more visible.

At $f=\SI{3.34}{\giga\hertz}$, the first step ($n=1$) is partially suppressed, but is recovered as frequency increases. The change in the oscillatory pattern is also clearly visible: at low frequency, fast oscillations follow a long region in which step $n=0$ (supercurrent) is present. As $f$ is increased, oscillations start earlier and with a larger period. These features are accounted for by the RSJ model presented in section \ref{RSJmodel}. More interestingly, half-integer subharmonic steps ($n=1/2,3/2,5/2...$) become clearly visible on the right panels (for $f\geq\SI{6.2}{\giga\hertz}$).

As discussed in ref.\onlinecite{Valizadeh2008}, it can be reproduced from the RSJ equations by adding a capacitive shunt in the circuit (RCSJ model \cite{McCumber1968}). Another possible mechanism is given by the presence of higher order harmonics in the current-phase relation. As a $\sin\phi/2$ term can suppress the odd steps, a $\sin2\phi$ term generates subharmonic steps. While the geometric capacitance is believed to be negligible (see estimate in ref.\onlinecite{Oostinga2013}), a non-sinusoidal current phase relation has clearly been established recently in our junctions \cite{Sochnikov2014}.

\begin{figure}[h!]
\centering\includegraphics[width=0.6\textwidth]{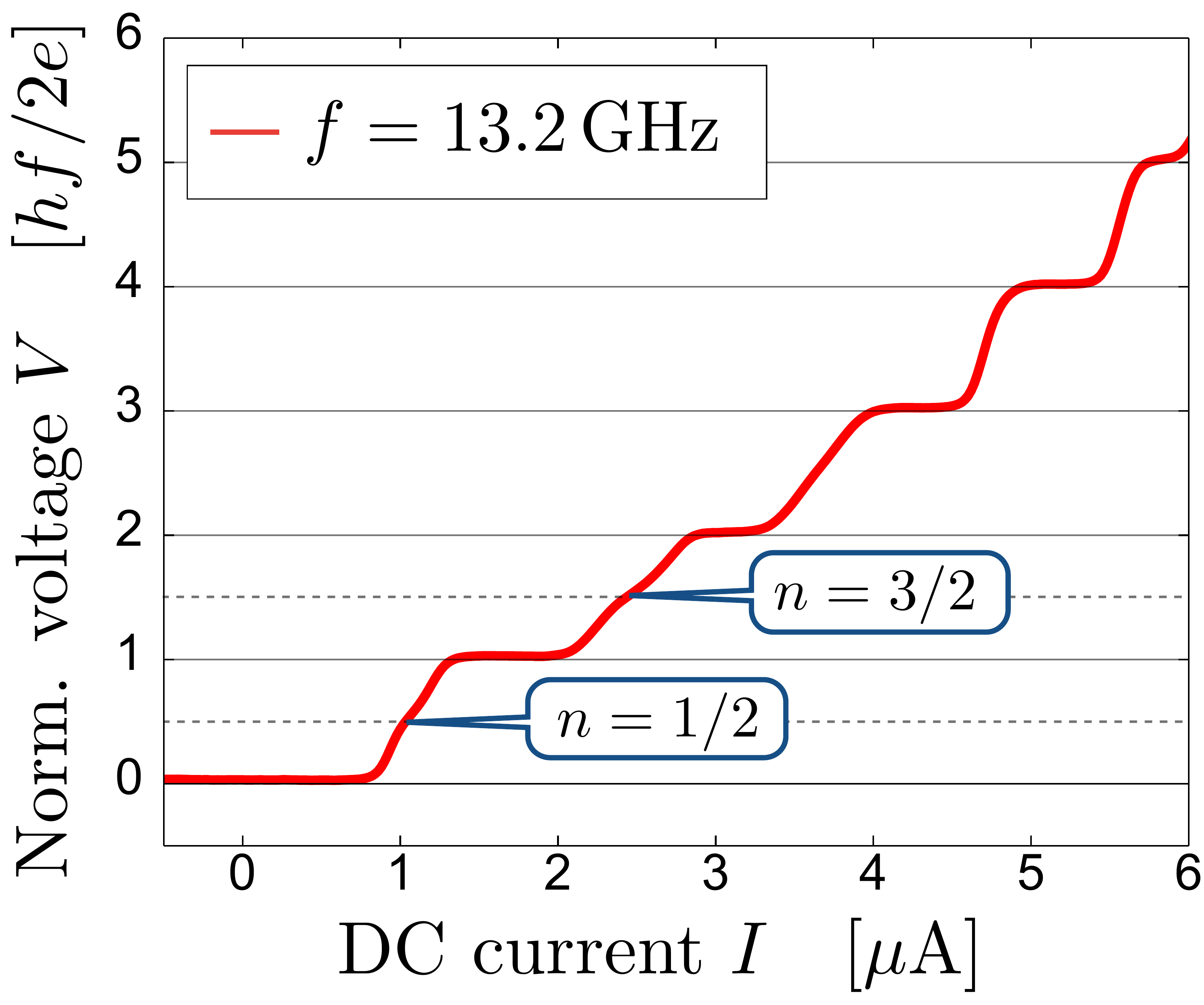}
\caption{{\bf Subharmonic steps -} We present an $I$-$V$ curve obtained under rf irradiation at $f=\SI{13.2}{\giga\hertz}$. The subharmonic structure (clearly visible in the differential resistance $dV/dI$ in Fig.\ref{Fig:SOMShapiroRock400nm}) is related to the appearance of subharmonic steps with index $n=1/2,3/2...$. Here in particular, one observes the steps $n=1/2$ and $n=3/2$.} \label{Fig:SOMShapiroSubharm}
\end{figure}

For the sake of completeness, we present in Fig.\ref{Fig:SOMShapiroSubharm} a dataset measured at $f=\SI{13.2}{\giga\hertz}$, on the junction with a length $L=\SI{150}{\nano\meter}$ presented in main text. The $I$-$V$ curve establishes the presence of two weak half-integer steps at $n=1/2$ and $3/2$.

\subsection{Summary of measured devices}
\label{Summary}
In the following table, we summarize the parameters of 6 devices for which a complete set of data has been collected. Parameters are as follows : length $L$, critical current $I_c$, normal state resistance $R_n$, Josephson frequency $f_J=\frac{2eR_n I_c}{h}$, crossover frequency $f_{4\pi}$, $4\pi$-periodic contribution to the supercurrent $I_{4\pi}$, induced gap $\Delta_i$, maximum current per mode $i_0=\frac{e\Delta}{\hbar}$, and finally number of $4\pi$-periodic modes. Starred ($^*$) values indicate devices for which the temperature dependence of the critical current has not been measured precisely, so that $\Delta_i$ is only estimated from the value of $I_c$ at base temperature, with large error bars. Despite a wide range of parameters explored, all junctions yield an estimated number of $4\pi$-periodic modes between 1 and 3. $^\dagger$ indicates the critical current of the junction with $L=\SI{150}{\nano\meter}$ presented in the main text at the time when the Shapiro steps have been measured. A previous cool-down of the sample yielded $I_c=\SI{5.6}{\micro\ampere}$ as presented in the main text (Fig.2). Aging of the sample has been observed in several samples and could explain this discrepancy.

\begin{center}
  \begin{tabular}{ | c | c | c | c | c | c | c | c | c | c |}
    \hline
     & $L$ (nm) & $I_c$ (\si{\micro\ampere})& $R_n$ (\si{\ohm}) & $f_J$ (GHz) & $f_{4\pi}$ (GHz) & $I_{4\pi}$ (nA) & $\Delta_i$ (meV)& $i_0$ (nA) &$N_{4pi}$ \\ \hline
  A & 150 & 3.3$^\dagger$ & 33 & 53 & 4.5-5 & 250-300 & 0.35 & 90 & 3 \\ \hline
  B & 400 & 0.29 & 158 & 22 & 4 & 55 & 0.1-0.15 & 25-40 & 2 \\ \hline
  C & 600 & 0.44 & 165 & 35 & 4  & 50 & 0.1-0.15& 25-40 & 1-2 \\ \hline
  D & 150 & 1.5 & 82  & 59 & 3-4  & 75-100  & 0.2-0.25$^*$& 50-65 & 1-2 \\ \hline
  E & 200 & 4.4 & 52  & 110 & 4-5  & 160-200  & 0.4-0.5$^*$& 100-130 & 1-2 \\ \hline
  F & 200 & 5.2 & 56  & 138 & 4-5  & 150-190  & 0.4-0.5$^*$& 100-130 & 1-2 \\ \hline
  \end{tabular}
\end{center}

\subsection{Weak reduction of the $n=3$ step}
\label{ThirdStep}
\begin{figure}[h!]
\centering\includegraphics[width=1\textwidth]{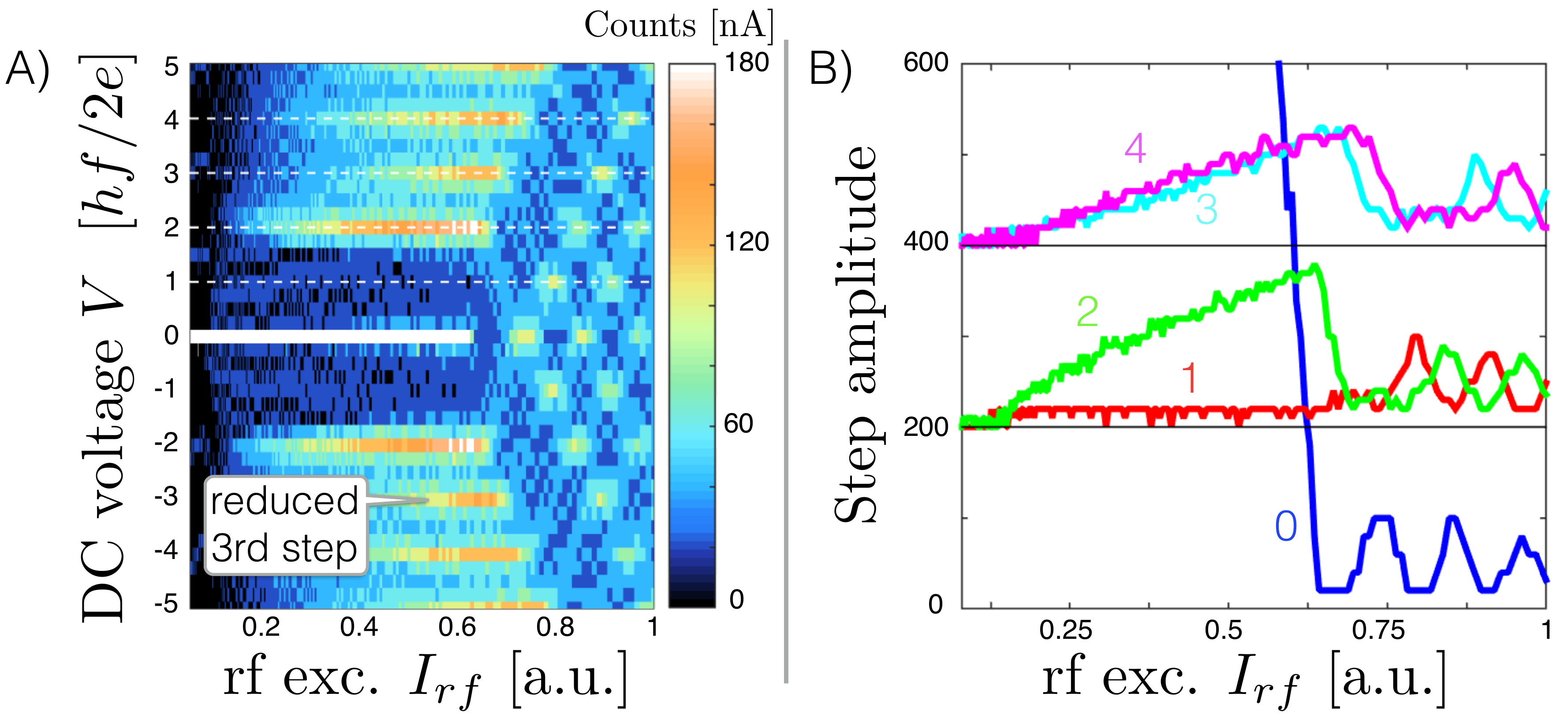}
\caption{{\bf Weak reduction of the $n=3$ step -} (A) The differential resistance $dV/dI$ is plotted as a colorscale, as a function of the dc and rf current drives $I$ and $I_{rf}$ for a frequency of $f=\SI{3}{\giga\hertz}$. A weak suppression of the third step ($n=3$) is observed for low rf currents. (B) The step amplitudes are extracted by binning voltages (see main text) and presented for steps 0 to 4 (with an offset of 0.03 for $n=1,2$ and 0.6 for $n=3,4$). As commonly observed in our samples, the first step (red line) if fully suppressed before the crossing point. More interestingly, the amplitude of the third step (cyan line) is reduced as compared with our usual observations. In particular, it remains smaller than the amplitude of the fourth step (presented in magenta)} \label{Fig:SOMThirdStep}
\end{figure}

In Fig.\ref{Fig:SOMThirdStep}A, we present the colormap of the bin counts obtained on a 200 nm long junction (not presented in the main text), as a function of the dc and rf drives $I$ and $I_{rf}$. The measurement is taken at $f=\SI{3}{\giga\hertz}$ and \SI{800}{\milli\kelvin}. For this frequency, we observe the complete suppression of the first Shapiro step. Again, the first oscillation is strongly disturbed as detailed in the main text. More surprisingly, a weak but distinct suppression of the third step $n=3$ can be seen for low rf currents. It is emphasized by plotting the amplitude of each pair of steps (Fig.\ref{Fig:SOMThirdStep}B). First, as described in the main text, the amplitude of the first step (red line) is fully suppressed up to the crossing point where the $n=0$ region vanishes, while the second step (green line) is fully visible. Then, the amplitude of the $n=3$ step (cyan line) is smaller than the amplitude of the $n=4$ one (purple line) while the opposite is usually expected from RSJ models.

This could constitute evidence of a missing or suppressed third step, clearly observed only in one sample. However, other irregularities in the higher order steps are sometimes visible (see Fig.3 in the main article), so that this evidence has to be considered carefully.

\subsection{Hysteresis, bias instability and Shapiro steps}

\begin{figure}[h!]
\centering\includegraphics[width=\textwidth]{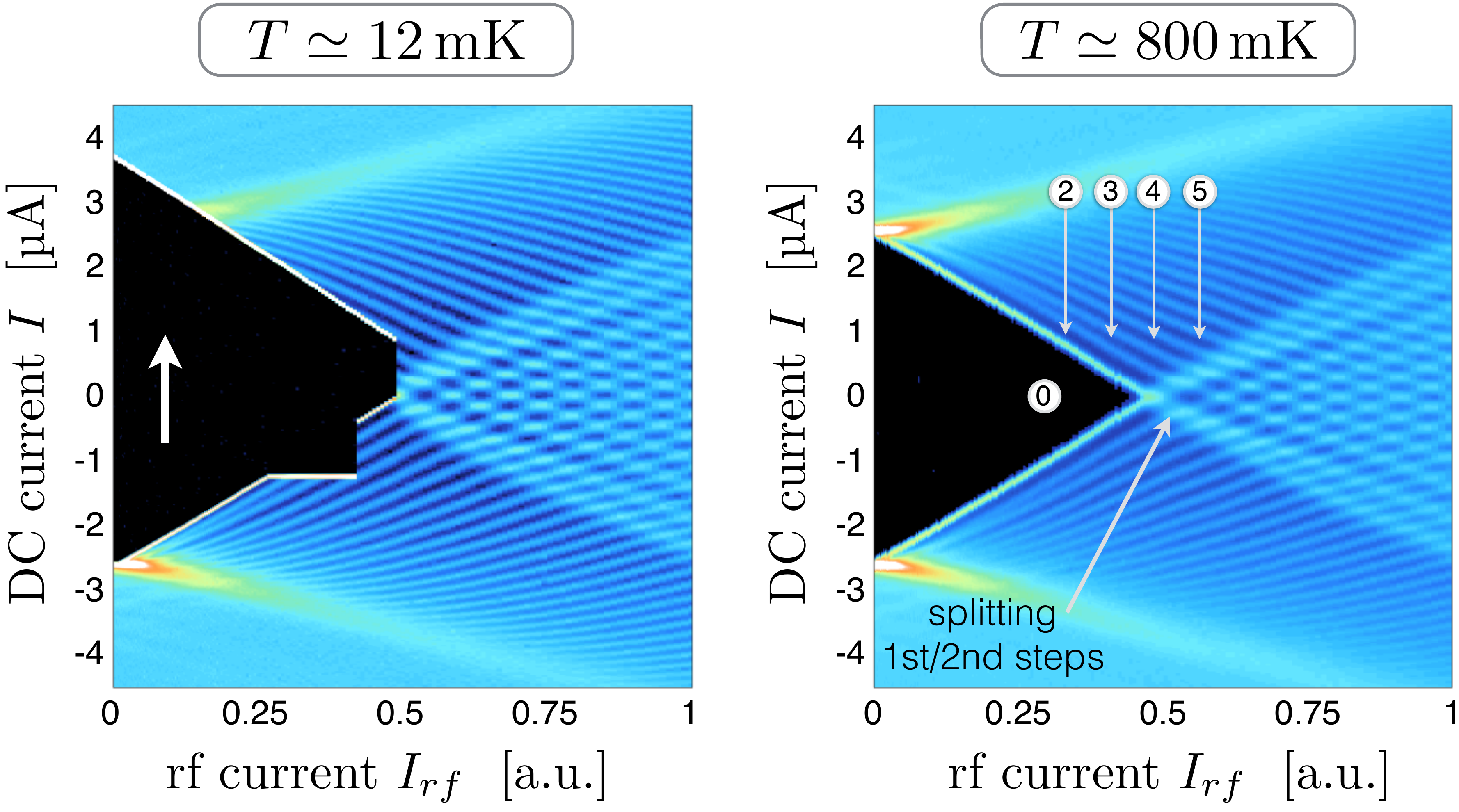}
\caption{{\bf Effect of hysteresis on Shapiro steps -} The differential resistance $dV/dI$ is plotted as a colorscale, as a function of the dc and rf current drives $I$ and $I_{rf}$ for a frequency of $f=\SI{3}{\giga\hertz}$, at the base temperature of the dilution refrigerator ($T=\SI{12}{\milli\kelvin}$) and at $T=\SI{800}{\milli\kelvin}$. The asymmetry in the figure signals a strong hysteretic behavior in the $I$-$V$ curve (see main text, Fig.2). The wide white arrow (left panel) symbolizes the sweep direction of the dc bias current $I$. In the hysteretic region, low index $n$ Shapiro steps are not visible. However, for high rf power, hysteresis vanishes and Shapiro steps reappear. The typical splitting of the $n=1/n=2$ steps is in particular still observable. In contrast, all steps (apart from the missing $n=1$) are visible at $T=\SI{800}{\milli\kelvin}$, and no asymmetry is observed.} \label{Fig:SOMHysteresis}
\end{figure}

In Fig.\ref{Fig:SOMHysteresis}, we present two datasets measured on the same junction as the one presented in section \ref{ThirdStep} at two different temperatures. To emphasize the role of the control parameters (dc and rf currents $I$ and $I_{rf}$), we now plot the differential resistance $dV/dI$ as a colormap, as a function of these two parameters. On the left panel, measurements were obtained at $f=\SI{3}{\giga\hertz}$ at the base temperature of the dilution refrigerator ($T=\SI{12}{\milli\kelvin}$). On the right panel, measurements were obtained in the same conditions ($f=\SI{3}{\giga\hertz}$) except for the temperature, here set high enough to suppress the hysteresis ($T\simeq800 \si{\milli\kelvin}$). 

Shapiro steps are identified as black regions where $dV/dI\simeq0$, while blue lines between black regions emphasize transitions between the different plateaus. A simultaneous reading of the voltage $V$ gives access to the step index $n$ (a few of them are indicated directly on Fig.\ref{Fig:SOMHysteresis}).
The dc current is swept in the direction indicated by the white arrow (from negative towards positive bias). At $T=\SI{12}{\milli\kelvin}$, the very clear asymmetry at low rf excitations is a signature of the hysteresis observed in the $I$-$V$ curve presented in the main text. The bistable dynamics that leads to hysteresis prevents the development of the phase-locked dynamics responsible for the Shapiro steps. Consequently, the latter are missing in the hysteretic region.
In contrast, all steps are clearly visible at higher temperature ($T\simeq800 \si{\milli\kelvin}$), except for the $n=1$ step which is fully suppressed at $f=\SI{3}{\giga\hertz}$.

It is possible to make a clear distinction between missing steps due to hysteresis and the missing $n=1$ step attributed to a $4\pi$-periodic supercurrent. First, as can be seen on Fig.\ref{Fig:SOMHysteresis}, hysteretic switchings present vertical tangents (similar to ref.\onlinecite{Courtois2008}) while the doubled Shapiro step always exhibits a finite slope. Second, hysteresis is characterized by an asymmetry depending on the sweep directions, which is not the case of the missing step in our measurements. Third, the anomalous splitting of the $n=1/n=2$ steps beyond the pinch-off of the supercurrent (corresponding to the dark fringe in the bin counts, discussed in the main text) that accompanies a missing $n=1$ step remains visible at all temperature regardless of the presence of hysteresis for low rf powers. For these reasons, it appears clear that one can safely neglect hysteresis as the origin of the missing $n=1$ step. 
To avoid any problem, most measurements were performed at a temperature high enough to suppress the hysteresis (typically 450 to 800 \si{\milli\kelvin}).

\subsection{Shapiro steps on a shunted device}

Furthermore, it is in fact possible to rule out definitively bias instabilities as a possible mechanism for the missing $n=1$ step. By adding a shunt resistance in parallel with the junction, one can indeed suppress hysteresis\cite{Chauvin2005} and work in a configuration that approaches the voltage bias regime. To do so, we add a \SI{10}{\ohm} resistor in series with the junction, and shunt these two elements with a \SI{1}{\ohm} resistor. Thus, the current flowing through the junction is accessed by measuring the voltage across the \SI{10}{\ohm} resistor, together with the voltage across the junction.

\begin{figure}[h!]
\centering\includegraphics[width=\textwidth]{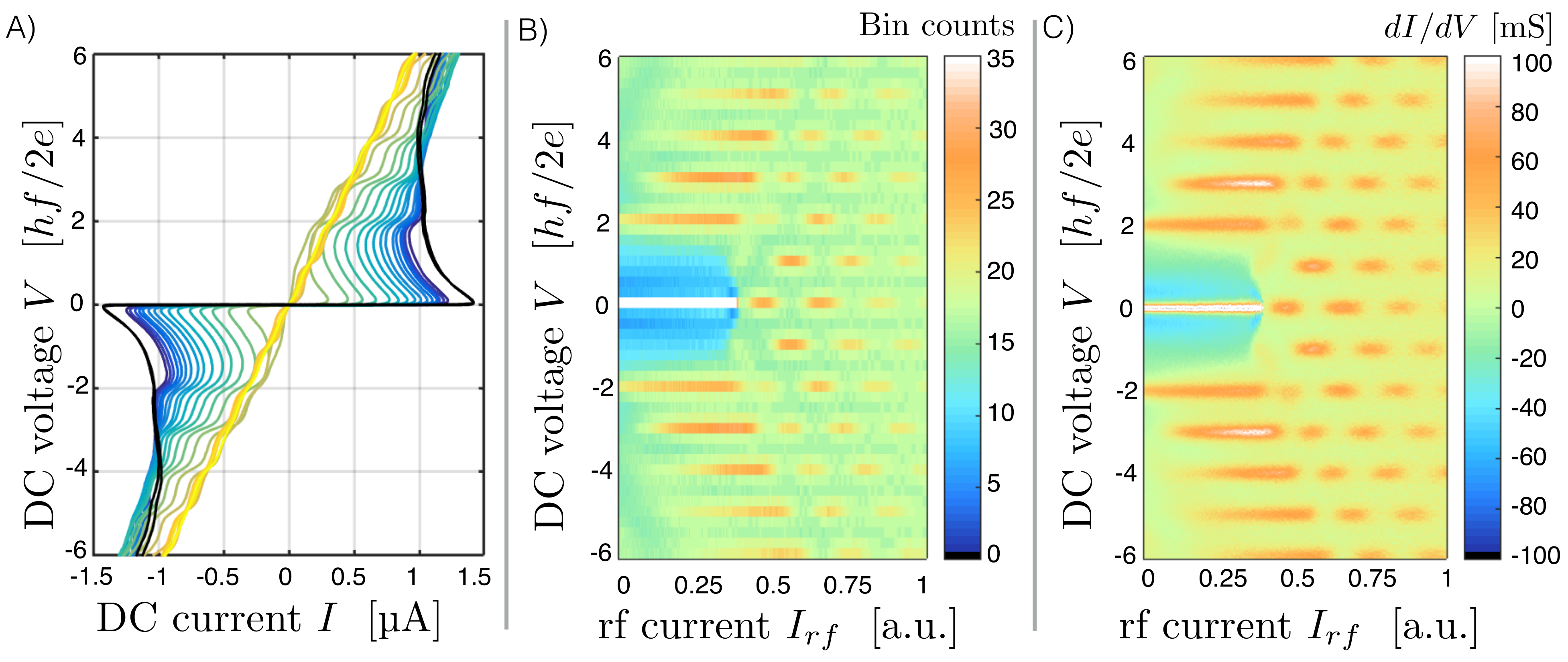}
\caption{{\bf Shapiro steps on a shunted device -} (A) $I-V$ curves in the shunted configuration, at temperature T=\SI{12}{\milli\kelvin}. A measurement in the absence of rf irradiation is shown as a black line. The low voltage regime with $dV/dI<0$ corresponds to an unstable region when a pure current bias is applied. The colored lines present data measured under irradiation at $f=\SI{4.2}{\giga\hertz}$. The rf current is increased progressively from indigo to yellow lines. (B) and (C) Histogram of the voltage and differential conductance $dI/dV$ as a function of DC voltage $V$ and rf current $I_{rf}$. As shown in the standard current bias configuration, a clear vanishing of the $n=1$ Shapiro step is observed.} \label{Fig:SOMVoltageBias}
\end{figure}

Results obtained at the base temperature of the fridge (12 mK) and with an rf excitation at \SI{4.2}{\giga\hertz} are shown in Fig.\ref{Fig:SOMVoltageBias}. First, the black solid line in Fig.\ref{Fig:SOMVoltageBias}A represents the $I-V$ curve in the absence of rf irradiation. At low voltages, one sees that $dV/dI<0$. In this region, a pure current bias generates bias instability and hysteresis, which is here suppressed by the shunt resistance. When the rf irradiation is switched on (colored plain lines), Shapiro steps become visible in the $I-V$ curves. As previously, the step $n=1$ is clearly suppressed at low rf excitation (blue lines). Increasing the rf drive (from blue to yellow lines), one sees the $n=1$ step is recovered at high drive amplitude as previously.

Fig.\ref{Fig:SOMVoltageBias} B and C present the voltage histograms and differential resistance $dI/dV$ as a function of the normalized DC voltage $V$ and rf current $I_{rf}$. Shapiro steps are visible as previously as maxima following horizontal lines. As for the measurements shown in the main text, the first step $n=1$ is fully suppressed up to the oscillating regions. Though the contrast is not as good, the "dark fringe" at finite voltage described in the main text is also visible.

\subsection{Magnetic field dependence}

\begin{figure}[h!]
\centering\includegraphics[width=0.6\textwidth]{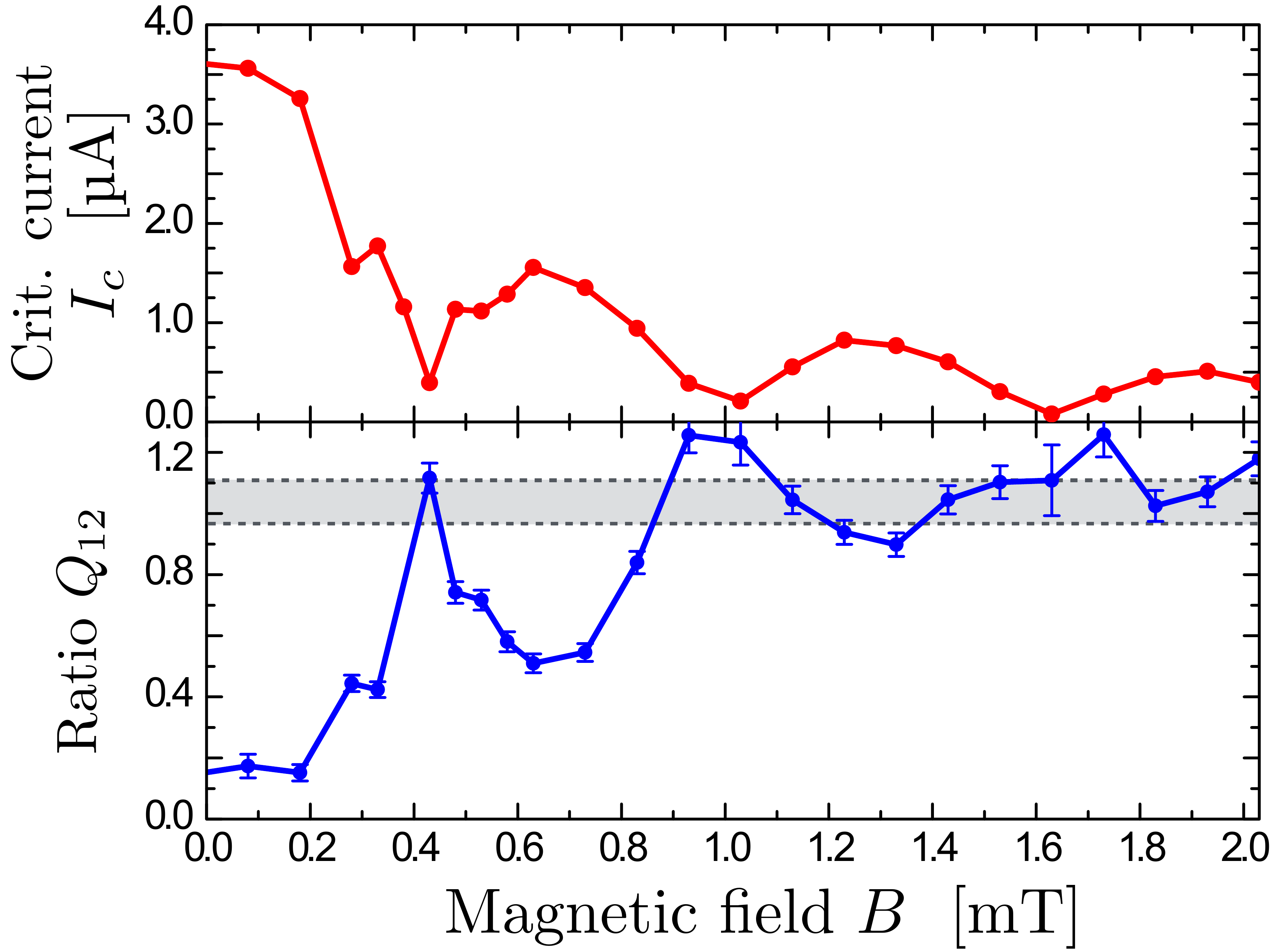}
\caption{ {\bf Magnetic field dependence -} (upper panel) Critical current $I_c$ as a function of the perpendicular magnetic field $B$. The typical diffraction pattern of the supercurrent is observed.
(lower panel) The ratio $Q_{12}$ is presented as a blue line, as a function of the perpendicular magnetic field $B$. It follows the same oscillations as the critical current $I_c$. The grey area between the dashed lines show the region in which the ratio $Q_{12}$ falls, as evaluated by a RSJ model.} \label{Fig:Bfield}
\end{figure}

Further investigations of the anomalous Shapiro response have been carried out in the presence of perpendicular-to-plane magnetic fields. First, when the $I$-$V$ curve is measured without rf irradiation, a Fraunhofer-like diffraction pattern of the critical current is observed (plotted as a red line in Fig.\ref{Fig:Bfield}, upper panel). In the junction presented here, the pattern is slightly distorted (probably due to flux-trapping in the magnet). The periodicity in the magnetic field has been evaluated from undistorted patterns in various other samples. It corresponds to a conventional periodicity as previously reported on similar samples\cite{Oostinga2013}, and as expected for ballistic systems with such aspect ratios\cite{Heida1998,Ledermann1999}. 

The observation of Shapiro steps in the presence of the magnetic field reveals that the anomalous doubling of the first step is only observed when the critical current is high enough, namely only at the center of the central lobe ($B\lesssim\SI{0.3}{\milli\tesla}$) and to a lesser extent at the centers of the second and third lobe. The ratio $Q_{12}$ of the first ($n=1$) and second ($n=2$) step widths (as introduced in the main text) is plotted as a function of the magnetic field in Fig.\ref{Fig:Bfield} (as a blue line, lower panel). The plot reproduces the variations observed in the critical current: the reduction of the first step is distinctly observed for high critical currents, yielding low values for $Q_{12}$, while $Q_{12}$ increases for low critical currents near the minima of the Fraunhofer pattern. This behavior can be understood by assuming a constant value of the $4\pi$-supercurrent fraction $I_{4\pi}/I_c$: the frequency $f_{4\pi}$ rapidly decreases when $I_c$ decreases, so that a correct observation of the anomalous step is not possible according to the criterion $f<f_{4\pi}$. 

One could expect that a small magnetic field induces a splitting of the Andreev bound states. This would consequently weaken the effect of Landau-Zener transitions (see section \ref{Section:LZT}) and allow us to exclude or confirm Landau-Zener transitions as responsible for the anomalous Shapiro response. Given the large $g$-factor in HgTe (around $g\simeq20$ for bulk HgTe), this effect could show up at relatively weak fields. To generate a splitting of $0.01\Delta_i$ that significantly alters Landau-Zener transitions, a typical magnetic field $B_0$ is required, with $g\mu_BB_0\simeq0.01\Delta_i$, yielding $B_0\simeq\SI{3.5}{\milli\tesla}$. The presence of a week magnetic field does not seem to suppress the $4\pi$-periodic contribution. It thus does not reveal any clear signature of energy splitting of the Andreev spectrum. This supports an interpretation based on the presence of a topological Andreev bound state. Indeed the latter is predicted to have no spin-degeneracy and therefore the level crossing at phase differences $\pi$, $3\pi$, etc. should persist even in the presence of a magnetic field on the TI.

\section{Additional results on graphene-based junctions}

\subsection{Typical properties of the graphene-based junctions}
In this section, we briefly present measurements of the Shapiro response of graphene-based devices. First, graphene flakes on hexagonal Boron-Nitride are fabricated by a van der Waals stacking method. Superconducting contacts are patterned by standard electron beam lithography and Niobium is deposited by magnetron sputtering. The geometry of the presented devices is similar to the one used for the HgTe devices, with the graphene weak link width ranging between \SI{1.5}{\micro\meter} and \SI{2.5}{\micro\meter}, and the length between \SI{200}{\nano\meter} and \SI{300}{\nano\meter}. 

\begin{figure}[h!]
\centering\includegraphics[width=0.9\textwidth]{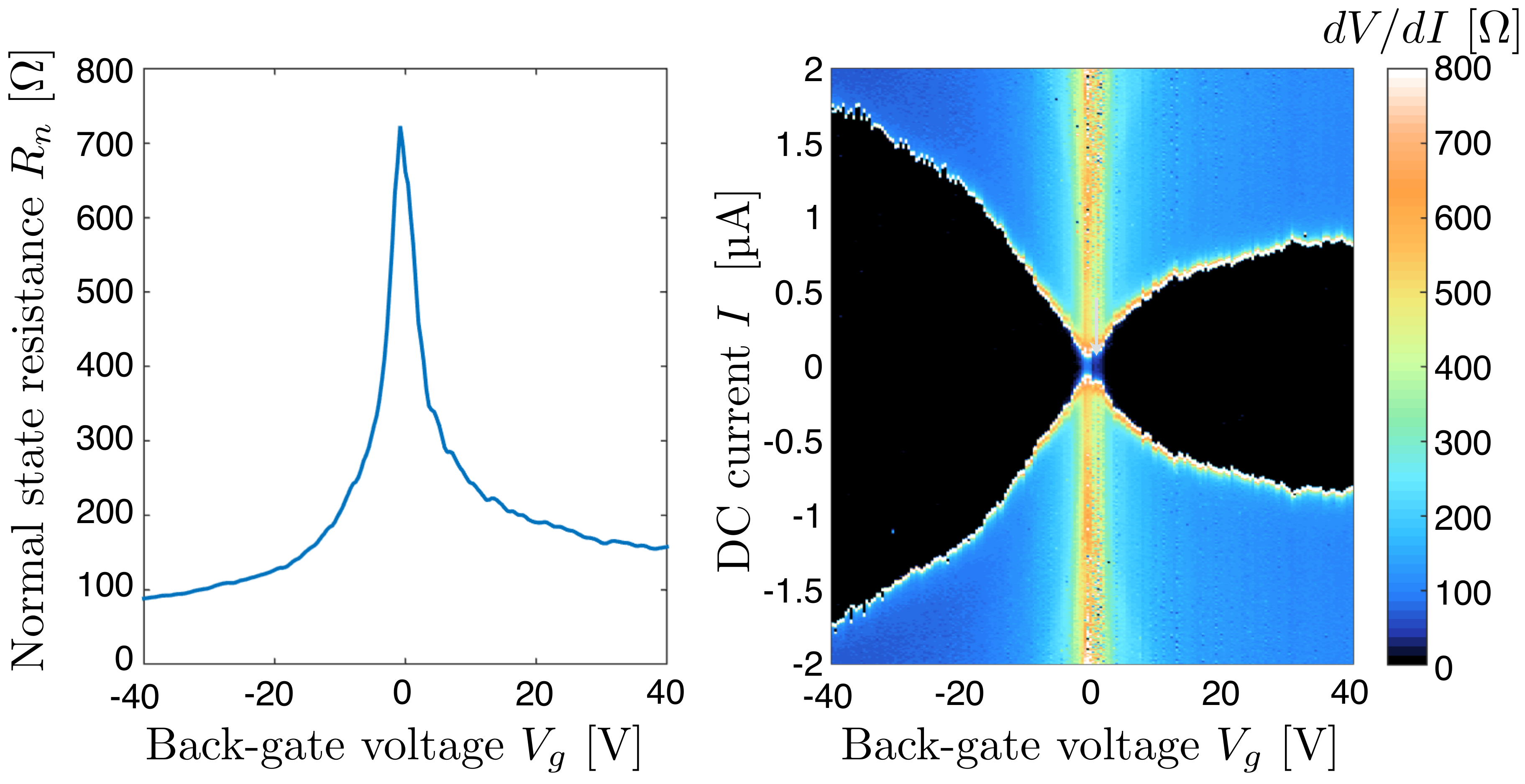}
\caption{{\bf DC characterization of Josephson junctions with graphene weak link -} (left) Normal state resistance $R_n$ of a graphene-based junction as a function of back-voltage $V_g$. The Dirac point is clearly indicated by a maximum around $V_g\simeq-1 V$. (right) Colormap of the differential resistance $dV/dI$ as a function of the back-gate voltage $V_g$ and dc drive current $I$. The supercurrent is visible as a black region ($dV/dI\simeq 0$) over the whole back-gate range, ranging from a few tens of \si{\nano\ampere} around the Dirac point up to \SI{1.8}{\micro\ampere} for large gate voltages.} \label{Fig:SOMGrapheneCharacterization}
\end{figure}

The graphene flake exhibit densities between \SI{-2.5e12}{\per\centi\meter\squared} and \SI{+2.5e12}{\per\centi\meter\squared}, which can be tuned by the means of a back-gate. In the left panel of Fig.\ref{Fig:SOMGrapheneCharacterization}, we present the extracted normal state resistance $R_n$ of the device as a function of the back-gate voltage $V_g$. The mobility approaches 3000 - \SI{8000}{\centi\meter\squared\per\volt\per\second}. Close to the Dirac point, this corresponds to a mean free path of around \SI{50}{\nano\meter}. Consequently, these devices are not in the ballistic limit but are relatively close to it.

In the right panel of Fig.\ref{Fig:SOMGrapheneCharacterization}, we present the differential resistance $dV/dI$ of one junction as a colormap, as a function of the back-gate voltage $V_g$ and dc drive current $I$. The supercurrent then appears as a dark region ($dV/dI\simeq 0$), while the normal state exhibits finite values of $dV/dI$. The devices show a supercurrent over the whole back-gate range, yielding supercurrents of a few tens of \si{\nano\ampere} around the Dirac point and up to \SI{1.8}{\micro\ampere} for large gate voltages.

\subsection{Shapiro response of the graphene-based junctions}

\begin{figure}[h!]
\centering\includegraphics[width=0.9\textwidth]{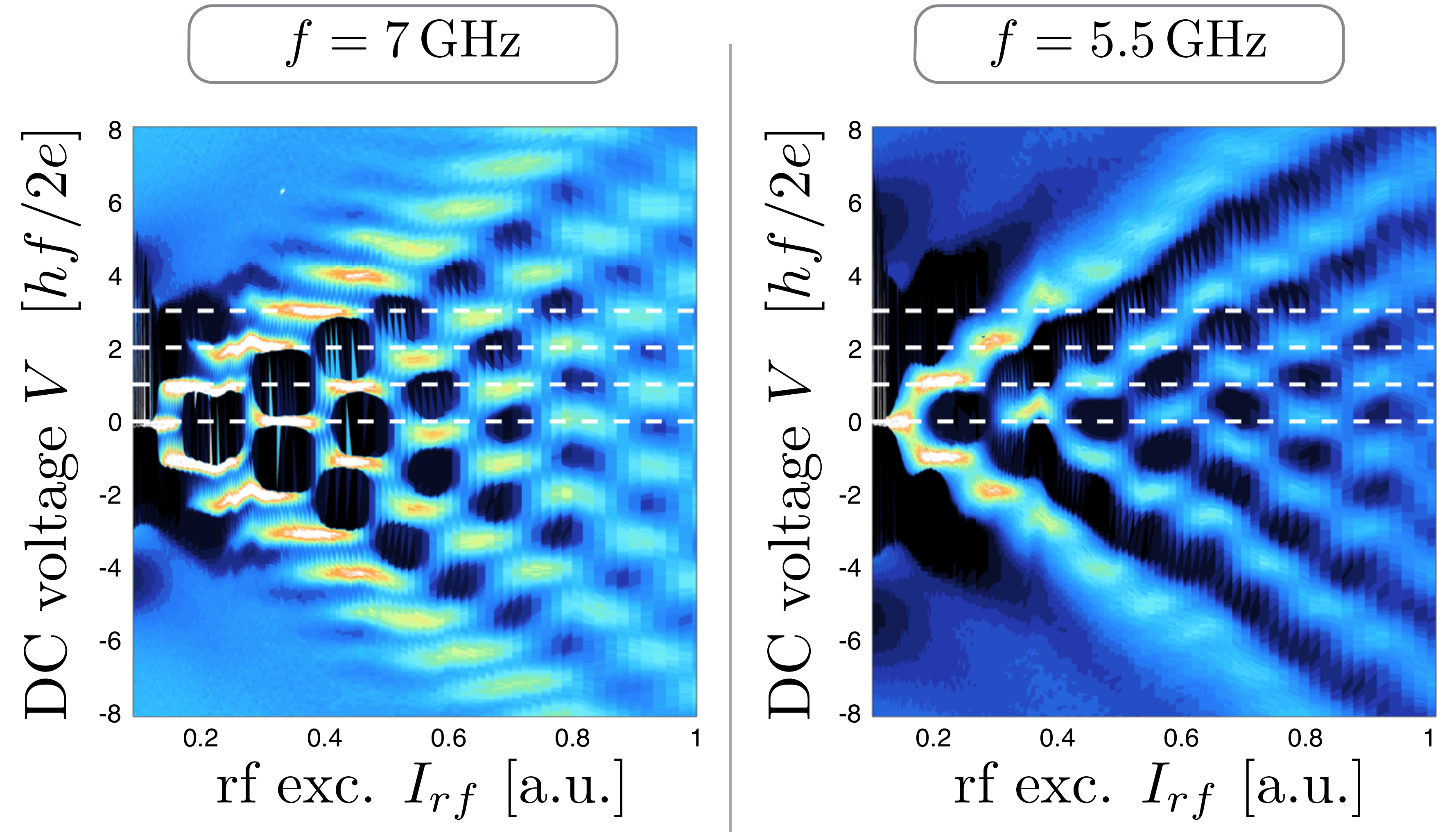}
\caption{{\bf Shapiro response in graphene-based junctions -} The differential conductance $dI/dV$ is plotted as a colorscale, as a function of the dc voltage $V$ and rf current drives $I_{rf}$  for two different frequencies $f=\SI{7}{\giga\hertz}$ and \SI{5.5}{\giga\hertz}. The data has been taken at the base temperature of the dilution refrigerator ($T=\SI{25}{\milli\kelvin}$), close to the Dirac point. All Shapiro steps are visible, up to $|n|>12$. } \label{Fig:SOMShapiroGraphene}
\end{figure}

In this section we present two typical sets of data taken on the graphene-based junctions (see Fig.\ref{Fig:SOMShapiroGraphene}). The differential conductance $dI/dV$ is plotted as a colorscale, as a function of the dc voltage $V$ and rf current drive $I_{rf}$ for two frequencies ($f=\SI{7}{\giga\hertz}$ and \SI{5.5}{\giga\hertz}). Despite fluctuations in the measurements (originating in fluctuations of the offset of one of our amplifiers), clear Shapiro steps are visible at the expected voltages $V_n=nhf/2e$. As frequency is lowered, the steps become less discernible but no sign of a vanishing $n=1$ step is observed in these sets. 

Measurements have been carried out for different values of the back-gate voltage (Dirac point, high $n$ or $p$ doping) and down to \SI{4}{\giga\hertz} (for which the steps are hardly visible) but no sign of a missing $n=1$ step has ever been spotted. Since no $4\pi$-periodic modes are expected in the graphene junction, the crossover frequency is expected to be  $f_{4\pi}=0$ in this system. These measurements establish an experimental upper boundary $f_{4\pi}\ll \SI{4}{\giga\hertz}$. Given the characteristic Josephson frequency of the device $f_J\simeq\SI{72}{\giga\hertz}$, this sets an upper boundary of $I_{4\pi}\ll \SI{75}{\nano\ampere}$.

\section{Elements of modeling}

\subsection{Simulations using RSJ equations}
\label{RSJmodel}
Simulations based on a standard RSJ model\cite{Russer1972} have been carried out to compare our results with a simple and well-understood model. In these simulations the junction is modeled together with a resistive shunt to capture the impedance of the environment (that plays an essential role in the dynamics of the junction). In this framework, the total current through the system $I$ can be written as the sum $I=I_R+I_S$ where $I_R=\frac{V}{R_n}$ is the current through the resistor $R_n$ and $I_S$ the supercurrent through the junction\footnote{As mentioned previously, the geometrical capacitance is small, and we neglect it here \cite{Oostinga2013}.}. Combining the first Josephson equation $d\phi/dt=2eV/\hbar$, and the current bias $I=I_{dc}+I_{rf}\sin 2\pi ft$ obtains a first order ordinary differential equation:
\begin{equation}
\frac{\hbar}{2e R_n}\dot\phi+I_S(\phi)=I_{dc}+I_{rf}\sin 2\pi ft\label{Eq:RSJ}
\end{equation}
The rf excitation is here represented as a current $I_{rf}$ instead of a voltage $V_{rf}$ in agreement with most of the literature on Shapiro steps. It assumes that the characteristic field impedance of the radiation field is high compared with the junction impedance \cite{Russer1972}. In our case (low capacitance), the typical impedance of the junction is given by its resistance (typically between 30 to \SI{150}{\ohm}), smaller than the free space impedance (in which RF propagates before the junction) given by $c\mu_0=1/c\epsilon_0\simeq\SI{376}{Ohm}$. This approximation is probably a bit crude, but it is further justified by the overall agreement of the effect of frequency on the Shapiro response (Fig. 3 of the main text and Fig. \ref{Fig:SOMShapiroRock400nm} and Fig.\ref{Fig:SOMLandauZener}), which is a characteristic feature of the current bias model developed by Russer \cite{Russer1972}. We simulate the results of this equation using a simple Runge-Kutta algorithm (RK4) to obtain the $I$-$V$ curve, realize a binning of the voltage $V$ and finally compute the ratios $Q_{12},Q_{34}$ for comparison with experimental results.

\begin{figure}[h!]
\centering\includegraphics[width=\textwidth]{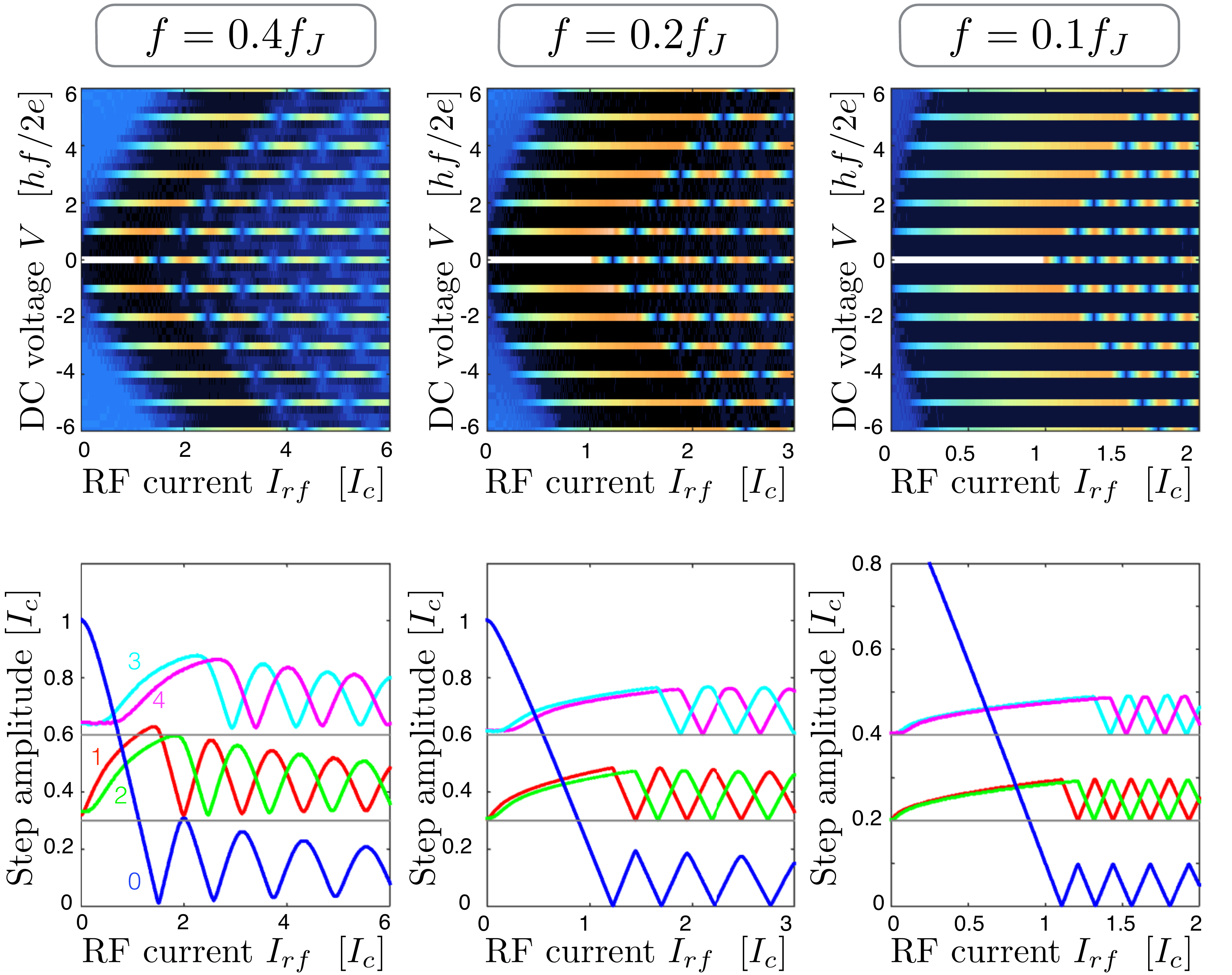}
\caption{{\bf Simulated Shapiro response of a conventional Josephson junction - } In the framework of the RSJ model (Eq.\ref{Eq:RSJ}), we obtain colormaps of the bin counts as a function of the voltage bins (in normalized units) and rf current drive $I_{rf}$. For a high frequency ($f=0.4 f_J$) the results obtained resembles that of Fig.3C in the main text. All steps are clearly visible and appear one by one as $I_{rf}$ increases. As frequency $f$ is decreased, the pinch-off of the supercurrent is moved to higher powers. One also observes a decreasing period in the oscillating pattern.} \label{Fig:SOMLandauZener}
\end{figure}

The results shown here are obtained with a $2\pi$-periodic sinusoidal current-phase relation $I_S(\phi)=I_c\sin\phi$. The results obtained by Russer\cite{Russer1972} are reproduced. The global visual agreement with our measurements is good, except the missing first step not accounted for in the standard RSJ model. In particular, the increasing frequency of oscillations (with respect to the rf current needed for the pinch-off of the supercurrent) is clearly observed.

In the lower panels, the step amplitudes for $n=0,1,2,3,4$ are plotted as a function of the rf current amplitude $I_{rf}$. Again, one observes the effect of the excitation frequency $f$. As $f$ decreases, the width of the first lobe gets much larger than the widths of the other lobes. Moreover, these graphs also show that the step amplitudes (hence their visibility) decrease (with respect to the critical current). This phenomenon, clearly observable in our measurements (lower panels of Fig.3, main text), limits our measurements to $f\geq\SI{2}{\giga\hertz}$.

\subsection{Extended RSJ model with a $4\pi$-periodic contribution:}

Simulations have also been carried out when adding a $4\pi$-periodic contribution following Dominguez {\it et al.}\cite{Dominguez2012}. We present in this section our results : the appearance of the doubled step is qualitatively well described as in the previous reference, but a quantitative agreement has not been obtained yet. The results are presented below in the following manner. First, we show that the addition of a $4\pi$-periodic contribution $I_{4\pi} \sin\phi/2$ to a sinusoidal $2\pi$-periodic current-phase relation (CPR) $I_{2\pi} \sin\phi$ is responsible for the disappearance of all odd steps at low frequency $f<f_{4\pi}$, in a comparable way to what we experimentally observe on the first step $n=1$. Then the marginal effect of the $2\pi$-periodic CPR is illustrated by comparing $I$-$V$ curves in the presence of rf irradiation for a few different CPRs. No generic $2\pi$-periodic CPR is found to show missing odd steps, while the addition of a small $4\pi$-periodic term enforces the disappearance of odd steps regardless of the $2\pi$-periodic term.

\paragraph{Effect of frequency on a $2\pi+4\pi$-periodic supercurrent:}

\begin{figure}[h!]
\centering\includegraphics[width=1\textwidth]{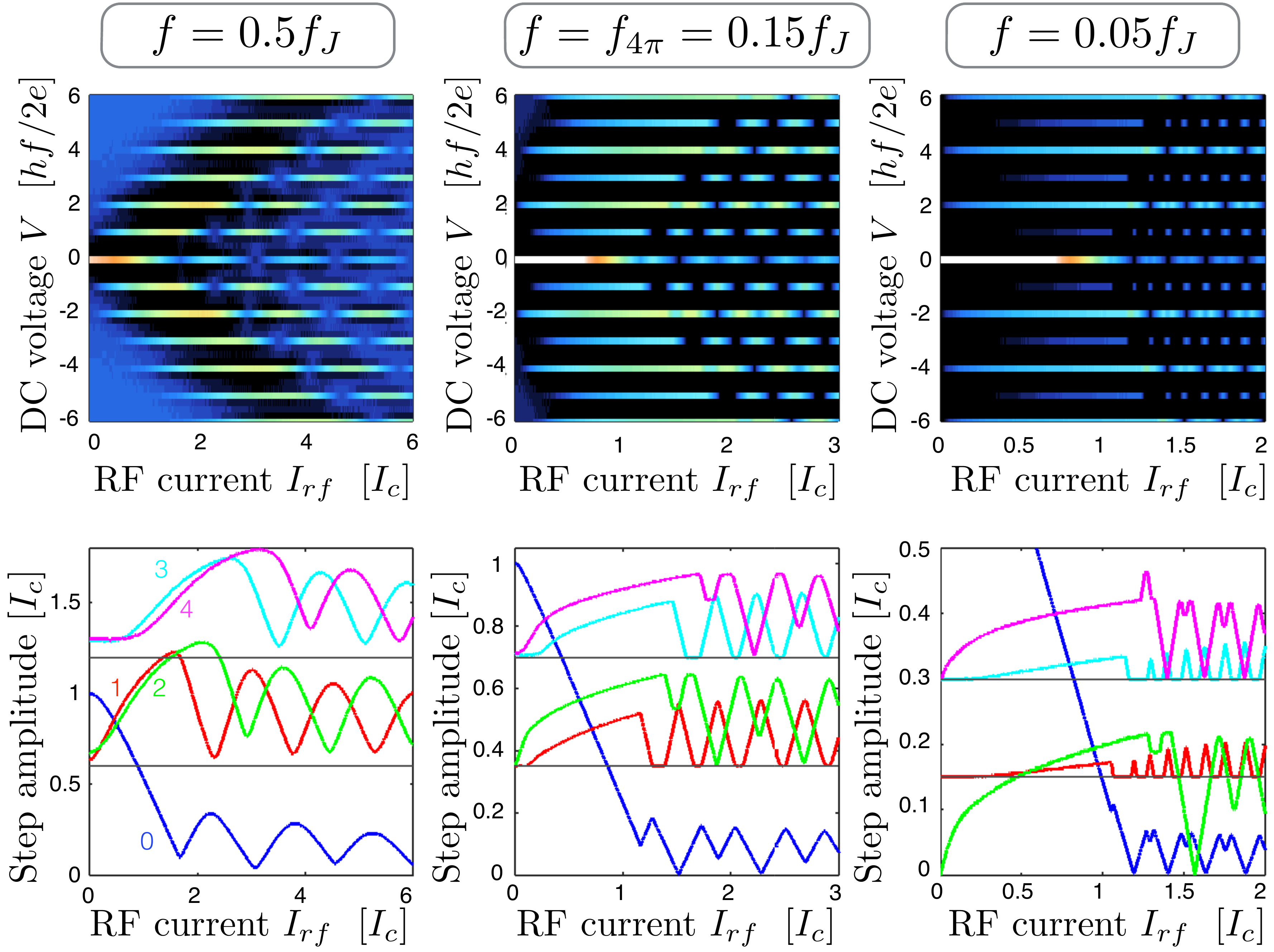}
\caption{{\bf Simulated Shapiro response with a $4\pi$-periodic term -} (upper panels) Colormaps of the bin counts as a function of the voltage (in normalized units) and rf current drive $I_{rf}$, in the presence of a small $4\pi$-periodic contribution (such that $f_{4\pi}=0.15 f_J$). (lower panels) Amplitudes of step $n=0$ to $n=4$, as a function of the rf current drive $I_{rf}$. At high frequency ($f=0.5 f_J>f_{4\pi}$), all steps are visible, and the result is very similar to  the one obtained without $4\pi$-periodic modes. As the frequency is decreased ($f=f_{4\pi}=0.15 f_J$ and $f=0.05f_J$), the amplitude of the odd steps decreases, in particular the $n=1$ step which is completely suppressed at low power. Simultaneously, the oscillatory pattern (high power) develops anomalies, with very pronounced first minima on the odd steps.} \label{Fig:SOMSimusRSJ4pi}
\end{figure}

Using Eq.(\ref{Eq:RSJ}), we simulate the effect of a small $4\pi$-periodic contribution $I_{4\pi} \sin\phi/2$. In Fig.\ref{Fig:SOMSimusRSJ4pi}, we show the effect of frequency on a CPR of the form $I_{4\pi} \sin\phi/2+ I_{2\pi} \sin\phi$ by plotting for different frequencies the histograms of the voltage as function of the rf current $I_{rf}$ (upper panels) and the amplitudes of steps $n=0$ to $n=4$ (lower panels). The parameters are $I_{2\pi}=1, I_{4\pi}=0.15$, so that $f_{4\pi}=0.15 f_J$. For a high frequency $f>f_{4\pi}$ (left, $f=0.5 f_J$), all steps are visible, and the 2D plot is similar to the one obtained without $4\pi$-periodic contribution. As frequency is lowered (center, $f=f_{4\pi}=0.15 f_J$, and right, $f=0.05f_J$), the amplitude of odd steps progressively decreases, and these odd steps are completely suppressed at low power.
The simulated behavior of the $n=1$ step is very similar to the one experimentally observed, and thus reinforces our interpretation. Though the crossover is not very sharp, it happens for $f=f_{4\pi}$ ($=0.15 f_J$ in the case of Fig.\ref{Fig:SOMSimusRSJ4pi}). One also observes that the oscillatory pattern (at high power) is also progressively modified from a $2\pi$- to a $4\pi$-dominated pattern. In particular, odd steps show a very pronounced first minimum. The dark fringe we experimentally observe in the oscillatory pattern is understood as the result of the progressive towards a pattern with a halved number of oscillations, thus yielding progressively suppressed lobes.

\paragraph{Effect of the CPR}

\begin{figure}[h!]
\centering\includegraphics[width=0.8\textwidth]{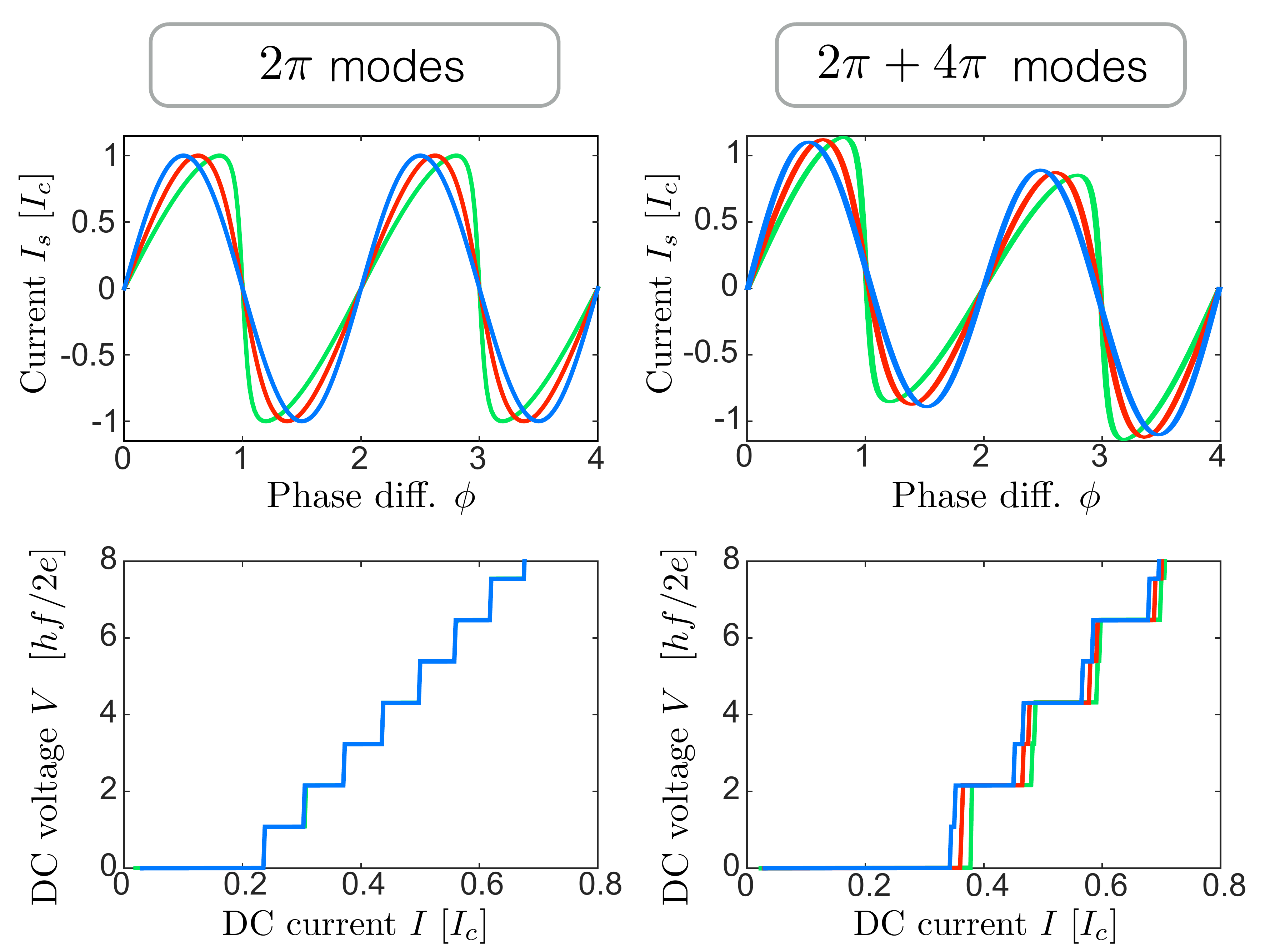}
\caption{{\bf Effect of the CPR -} (upper panels) Investigated current-phase relations: in blue, the sinusoidal 2π-periodic supercurrent $I_{2\pi} \sin\phi$, in red and green, single-mode-like supercurrent with transmission $\tau=0.8$ and 0.99 respectively. On the right side, an additional $4\pi$-periodic contribution is added, with amplitude $I_{4\pi}=0.15 I_{2\pi}$. (lower panels) $I$-$V$ curves in the presence of rf excitation. While all steps are visible in the absence of 4π-periodic modes, odd steps are missing or strongly suppressed when they are present. The exact CPR plays a small role in the crossover from the $4\pi$-dominated regime to the conventional one when frequency is increased.} \label{Fig:SOMEffectCPR}
\end{figure}

Though the presence of a small $4\pi$-periodic contribution $I_{4\pi} \sin\phi/2$ is found necessary to observe vanishing odd steps, the exact description of the $2\pi$-periodic supercurrent does not influence much the Shapiro response. To illustrate this finding, we focus on three different CPR. The first one is a standard $I_{2\pi} \sin\phi$ contribution (obtained for a tunnel junction for example), as in the previous paragraph. For the other two, we assume that the current is carried as a single mode of transmission $\tau=0.8$ and 0.99 respectively. The skewness of the CPR increases with transmission. The current carried by such a mode is then given by $\tau \sin\phi/\sqrt{1-\tau\sin^2\phi/2}$. We normalize all CPRs such that $I_c=I_{2\pi}=1$. Fig.\ref{Fig:SOMEffectCPR} shows in the upper panels the CPR and in the lower panels an $I$-$V$ curve in the presence of rf irradiation, for a low frequency $f=0.05 f_J=f_{4\pi}/3$. On the left side, the CPR contains only this $2\pi$-periodic supercurrent, and the obtained $I$-$V$ curves exhibits all integer steps and do not depend on the exact CPR. When a $4\pi$-periodic supercurrent is added, all odd steps are suppressed or reduced. The CPR matters marginally, and only seem to control the exact detail of the transition from the $2\pi$- to the $4\pi$-dominated regime : the transition is slightly faster for a skewed CPR.

\subsection{Evaluation of Landau-Zener probabilities}
\label{Section:LZT}
\begin{figure}[h!]
\centering\includegraphics[width=0.6\textwidth]{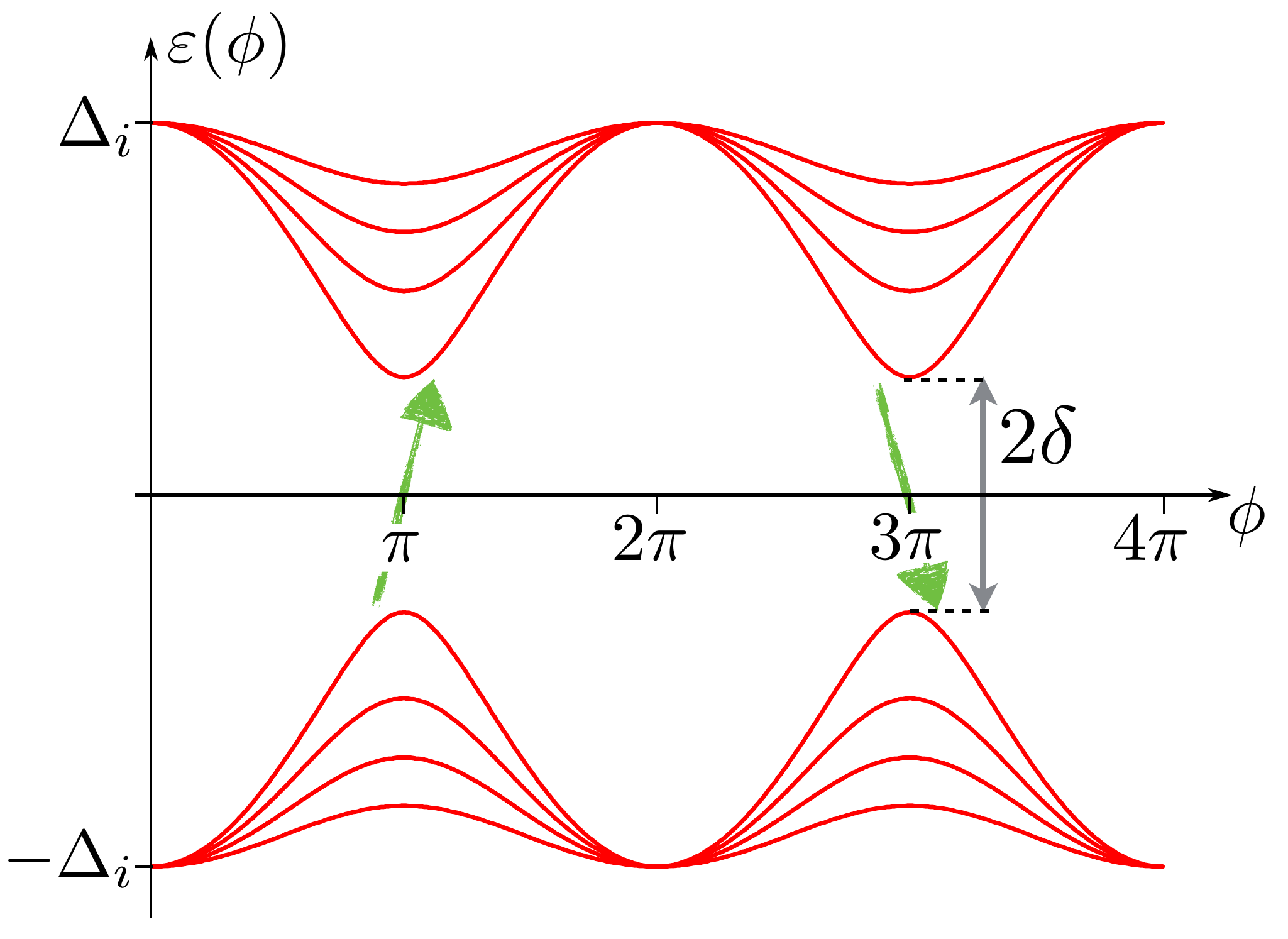}
\caption{{\bf Effect of Landau-Zener transitions -} A typical spectrum of gapped $2\pi$-periodic Andreev levels is presented as a function of the superconducting phase difference $\phi$. In the presence of Landau-Zener transitions at the anticrossings (represented as green arrows for $\phi=\pi, 3\pi,...$), some energy levels with small gap $\delta$ (high transmission) could contribute as a $4\pi$-periodic component in the supercurrent.} \label{Fig:SOMAndreevSpectrumLZT}
\end{figure}

If some Andreev bound states have very high transmission (or small energy gap $\delta$, see Fig.\ref{Fig:SOMAndreevSpectrumLZT}), non-adiabatic Landau-Zener processes may be responsible for an additional $4\pi$-periodic contribution to the supercurrent. Namely, some gapped $2\pi$-periodic Andreev levels would behave as effective $4\pi$-periodic modes, in the absence of true $4\pi$-periodic modes.
To our knowledge, studies of the effect of Landau-Zener transitions on the dynamics of Josephson junctions are scarce \cite{Averin1995,Goffman2000, Chauvin2007, Dominguez2012, Sau2012} in general and systems with multiple Andreev levels have not been explored yet. We thus use a single mode approximation \cite{Dominguez2012}, and crudely assume that one level (with lowest gap $\delta$) has a predominant role in Landau-Zener processes. 

Following Dominguez {\it et al.} \cite{Dominguez2012}, we include in our simulations stochastic Landau-Zener transitions at the anticrossings, with a probability $P$, and partially reproduce their results. The exact motion of the phase difference is hard to picture, but the general trend can be understood on a heuristic basis. The phase will undergo shifts of $4\pi$ (per period of the drive) in the case of a Landau-Zener transition and only $2\pi$ in the absence of transitions. On average this results in a non-universal Shapiro step that is neither $hf/2e$ nor $hf/e$. These two values are recovered for $P=0$ and $P=1$ respectively. The results of our simulations are presented in Fig.\ref{Fig:SOMLandauZener}A, in which a close-up of the $n=2$ Shapiro step is presented, for different values of $P$.  As Dominguez {\it et al.}, we observe a splitting of the step for $P$ close to 1, with voltage plateaus that deviate from the quantized value $hf/e$ until the split steps are eventually less discernible for $P\lesssim 0.7$. Our experimental results do not show any splitting or deviation to the quantized value, with an accuracy of a few percents. This indicates that the probability must be $P>0.97$ ($P=1$ being equivalent to having a fully gapless mode). 

From the previous model, one can estimate the importance of the Landau-Zener transitions. First, we solve Eq.\ref{Eq:RSJ}, and obtain the phase $\phi(t)$ and its derivative  
$\dot\phi(t)\propto V(t)$ as a function of time $t$. The parameters are chosen such that the junction lies on the first Shapiro step ($V=\langle V(t)\rangle=\frac{hf}{2e}$). In that case, a high probability of Landau-Zener transitions would lead the junction to exhibit a doubled step. 

A typical plot of $\phi(t)$ and $V(t)$ is shown in Fig.\ref{Fig:SOMLandauZener}B, with $\phi(t)$ as a red line and $V(t)\propto\dot\phi(t)$ as a blue line for the following parameters: $I=0.5\, I_c, I_{rf}=0.8\, I_c, f=0.2\, f_J$. One first observes that the phase $\phi$ follows an anharmonic motion synchronized with the excitation drive at frequency $f$: during one period of duration $1/f$, the phase $\phi$ increases by $2\pi$, yielding an averaged voltage $V=hf/2e$ as expected for the first Shapiro step. Equivalently, one can calculate the average of $V(t)$ and obtain $V=\langle V(t)\rangle=\frac{hf}{2e}$.
Then, we access the time $t$ for which $\phi$ reaches the anticrossing (for $\phi(t)=3\pi$ for example) and read the derivative of the phase $\dot\phi\vert_{3\pi}$ at this point or equivalently the voltage $V\vert_{3\pi}$. Finally, the Landau-Zener probability $P$ is obtained from the following equation \cite{Dominguez2012}:
$$P=\exp\Big(-2\pi\frac{\delta^2}{\Delta_i\hbar\dot\phi}\Big)$$
This assumes a generic pair of Andreev bound state with energy $\varepsilon_\pm(\phi)=\pm\sqrt{\delta^2+\Delta_i^2\cos^2\phi/2}$.
\begin{figure}[h!]
\centering\includegraphics[width=\textwidth]{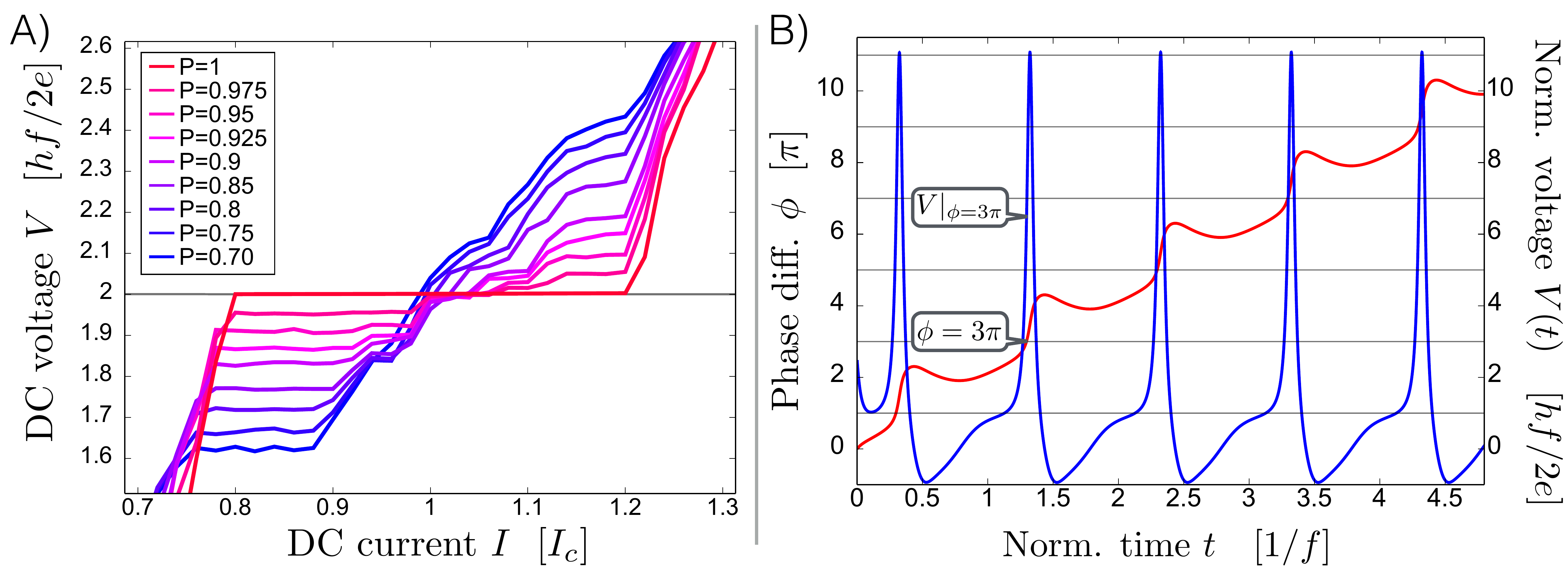}
\caption{{\bf Landau-Zener transitions in the RSJ model - } A) Evolution of the Shapiro step $n=2$: $I$-$V$ curves obtained for different values of the Landau-Zener probability $P$ are plotted, with a focus on the $n=2$ Shapiro step. When $P=1$, the second step ($n=2$) is fully developed and reaches the expected quantized value $hf/e$. For $P<1$, it progressively deteriorates: the voltage departs from the quantized value $hf/2e$, and the plateau becomes less visible. B) Time evolution of phase difference $\phi(t)$ and voltage $V(t)$: The phase difference $\phi(t)$ (red line, in units of $\pi$) and voltage $V(t)$ (blue line, in units $hf/2e$) across the junction is calculated via RSJ equations and plotted as a function of time $t$ in units of the rf period $1/f$. The estimation of the voltage $V\vert_{3\pi}$ at $\phi(t)=3\pi$ (anticrossing) enables a proper evaluation of the Landau-Zener transition probability.} \label{Fig:SOMLandauZener}
\end{figure}
For the graphs presented here, we obtain $V\vert_{3\pi}\simeq 6.4\, \frac{hf}{2e}$. This yields a probability $P>0.98$ if $\delta<\SI{6}{\micro\electronvolt}$. The most stringent constraint is obtained for the lowest frequencies accessible in the experiment (\SI{2}{\giga\hertz}). With the criterion $P>0.97$, we obtain $\delta\lesssim \SI{9}{\micro\electronvolt}$ for $\Delta_i=\SI{0.12}{\milli\electronvolt}$ ($L= 400$ and \SI{600}{\nano\meter}), and $\delta\lesssim \SI{18}{\micro\electronvolt}$ for $\Delta_i=\SI{0.35}{\milli\electronvolt}$ ($L= \SI{150}{\nano\meter}$). This corresponds to a transmission $\sqrt{1-(\delta/\Delta_i)^2}\geq0.994$ in both cases.

\clearpage

\bibliographystyle{unsrt}

\bibliography{BibShapiro.bib}

\begin{thebibliography}{10}

\bibitem{Fu2007}
L.~Fu, C.~Kane, and E.~Mele.
\newblock {Topological Insulators in Three Dimensions}.
\newblock {\em Physical Review Letters}, 98(10):106803, 2007.

\bibitem{Fu2008}
L.~Fu and C.~Kane.
\newblock {Superconducting Proximity Effect and Majorana Fermions at the
  Surface of a Topological Insulator}.
\newblock {\em Physical Review Letters}, 100(9):096407, 2008.

\bibitem{Olund2012}
C.T. Olund and E.~Zhao.
\newblock {Current-phase relation for Josephson effect through helical metal}.
\newblock {\em Physical Review B}, 86(21):214515, 2012.

\bibitem{Beenakker2013}
C.W.J. Beenakker.
\newblock {Search for Majorana Fermions in Superconductors}.
\newblock {\em Annual Review of Condensed Matter Physics}, 4(1):113--136, 2013.

\bibitem{Tkachov2013}
G.~Tkachov and E.M. Hankiewicz.
\newblock {Helical Andreev bound states and superconducting Klein tunneling in
  topological insulator Josephson junctions}.
\newblock {\em Physical Review B}, 88(7):075401, 2013.

\bibitem{Kitaev2001}
A.~Kitaev.
\newblock {Unpaired Majorana fermions in quantum wires}.
\newblock {\em Physics-Uspekhi}, 44:16, 2001.

\bibitem{Zhang2014}
F.~Zhang and C.L. Kane.
\newblock {Anomalous topological pumps and fractional Josephson effects}.
\newblock {\em Physical Review B}, 90(2):020501, 2014.

\bibitem{Veldhorst2012}
M.~Veldhorst, M.~Snelder, M.~Hoek, T.~Gang, V.K. Guduru, X.L. Wang, U.~Zeitler,
  W.G. van~der Wiel, A.A. Golubov, H.~Hilgenkamp, and A.~Brinkman.
\newblock {Josephson supercurrent through a topological insulator surface
  state.}
\newblock {\em Nature materials}, 11(5):417--21, 2012.

\bibitem{Oostinga2013}
J.B. Oostinga, L.~Maier, P.~Sch\"{u}ffelgen, D.~Knott, C.~Ames, C.~Br\"{u}ne,
  G.~Tkachov, H.~Buhmann, and L.W. Molenkamp.
\newblock {Josephson Supercurrent through the Topological Surface States of
  Strained Bulk HgTe}.
\newblock {\em Physical Review X}, 3(2):021007, 2013.

\bibitem{Maier2015}
L.~Maier, E.~Bocquillon, M.~Grimm, J.B. Oostinga, C.~Ames, C.~Gould,
  C.~Br\"{u}ne, H.~Buhmann, and L.W. Molenkamp.
\newblock {Phase-sensitive SQUIDs based on the 3D topological insulator HgTe}.
\newblock {\em Physica Scripta}, T164(1):014002, 2015.

\bibitem{Kurter2014}
C.~Kurter, A.D.K. Finck, P.~Ghaemi, Y.S. Hor, and D.J. {Van Harlingen}.
\newblock {Dynamical gate-tunable supercurrents in topological Josephson
  junctions}.
\newblock {\em Physical Review B}, 90(1):014501, 2014.

\bibitem{Galletti2014}
L.~Galletti, S.~Charpentier, M.~Iavarone, P.~Lucignano, D.~Massarotti,
  R.~Arpaia, Y.~Suzuki, K.~Kadowaki, T.~Bauch, A.~Tagliacozzo, F.~Tafuri, and
  F.~Lombardi.
\newblock {Influence of topological edge states on the properties of $\rm
  Bi_2Se_3/Al$ hybrid Josephson devices}.
\newblock {\em Physical Review B}, 89(13):134512, 2014.

\bibitem{Finck2014}
A.D.K. Finck, C.~Kurter, Y.S. Hor, and D.J. {Van Harlingen}.
\newblock {Phase Coherence and Andreev Reflection in Topological Insulator
  Devices}.
\newblock {\em Physical Review X}, 4(4):041022, 2014.

\bibitem{Kwon2004}
H.-J. Kwon, V.M. Yakovenko, and K.~Sengupta.
\newblock {Fractional ac Josephson effect in unconventional superconductors}.
\newblock {\em Low Temperature Physics}, 30(7):613, 2004.

\bibitem{SanJose2012}
P.~San-Jose, E.~Prada, and R.~Aguado.
\newblock {ac Josephson Effect in Finite-Length Nanowire Junctions with
  Majorana Modes}.
\newblock {\em Physical Review Letters}, 108(25):257001, 2012.

\bibitem{Pikulin2012}
D.I. Pikulin and Y.V. Nazarov.
\newblock {Phenomenology and dynamics of a Majorana Josephson junction}.
\newblock {\em Physical Review B}, 86(14):140504, 2012.

\bibitem{Badiane2013}
D.M. Badiane, L.I. Glazman, M.~Houzet, and J.S. Meyer.
\newblock {Ac Josephson effect in topological Josephson junctions}.
\newblock {\em Comptes Rendus Physique}, 14(9-10):840--856, 2013.

\bibitem{Shapiro1963}
S.~Shapiro.
\newblock {Josephson Currents in Superconducting Tunneling: The Effect of
  Microwaves and Other Observations}.
\newblock {\em Physical Review Letters}, 11(2):80--82, 1963.

\bibitem{Rokhinson2012}
L.P. Rokhinson, X.~Liu, and J.K. Furdyna.
\newblock {The fractional a.c. Josephson effect in a
  semiconductor/superconductor nanowire as a signature of Majorana particles}.
\newblock {\em Nature Physics}, 8(11):795--799, 2012.

\bibitem{Tanaka1997}
Y.~Tanaka and S.~Kashiwaya.
\newblock {Theory of Josephson effects in anisotropic superconductors}.
\newblock {\em Physical Review B}, 56(2):892--912, 1997.

\bibitem{Bruene2011}
C.~Br\"{u}ne, C.X. Liu, E.G. Novik, E.M. Hankiewicz, H.~Buhmann, Y.L. Chen,
  X.L. Qi, Z.X. Shen, S.C. Zhang, and L.W. Molenkamp.
\newblock {Quantum Hall Effect from the Topological Surface States of Strained
  Bulk HgTe}.
\newblock {\em Physical Review Letters}, 106(12):126803, 2011.

\bibitem{Bruene2014}
C.~Br\"{u}ne, C.~Thienel, M.~Stuiber, J.~B\"{o}ttcher, H.~Buhmann, E.G. Novik,
  C.-X. Liu, E.M. Hankiewicz, and L.W. Molenkamp.
\newblock {Dirac-screening stabilized surface-state transport in a topological
  insulator}.
\newblock {\em Physical Review X}, (14):041045, 2014.

\bibitem{Renne1974}
M.J. Renne and D.~Polder.
\newblock {Some analytical results for the resistively shunted Josephson
  junction}.
\newblock {\em Revue de Physique Appliqu\'{e}e}, 9(1):25--28, 1974.

\bibitem{Valizadeh2008}
A.~Valizadeh, M.R. Kolahchi, and J.P. Straley.
\newblock {On the Origin of Fractional Shapiro Steps in Systems of Josephson
  Junctions with Few Degrees of Freedom}.
\newblock {\em Journal of Nonlinear Mathematical Physics}, 15(sup3):407--416,
  2008.

\bibitem{Dominguez2012}
F.~Dom\'{\i}nguez, F.~Hassler, and G.~Platero.
\newblock {Dynamical detection of Majorana fermions in current-biased
  nanowires}.
\newblock {\em Physical Review B}, 86(14):140503, 2012.

\bibitem{Fu2007a}
L.~Fu and C.~Kane.
\newblock {Topological insulators with inversion symmetry}.
\newblock {\em Physical Review B}, 76(4):045302, 2007.

\bibitem{Courtois2008}
H.~Courtois, M.~Meschke, J.~Peltonen, and J.~Pekola.
\newblock {Origin of Hysteresis in a Proximity Josephson Junction}.
\newblock {\em Physical Review Letters}, 101(6):067002, 2008.

\bibitem{Blonder1982}
G.E. Blonder, M.~Tinkham, and T.M. Klapwijk.
\newblock {Transition from metallic to tunneling regimes in superconducting
  microconstrictions: Excess current, charge imbalance, and supercurrent
  conversion}.
\newblock {\em Physical Review B}, 25(7):4515--4532, 1982.

\bibitem{Klapwijk1982}
T.M. Klapwijk, G.E. Blonder, and M.~Tinkham.
\newblock {Explanation of subharmonic energy gap structure in superconducting
  contacts}.
\newblock {\em Physica B+C}, 109-110:1657--1664, 1982.

\bibitem{Scheer2001}
E.~Scheer, W.~Belzig, Y.~Naveh, M.~Devoret, D.~Esteve, and C.~Urbina.
\newblock {Proximity Effect and Multiple Andreev Reflections in Gold Atomic
  Contacts}.
\newblock {\em Physical Review Letters}, 86(2):284--287, 2001.

\bibitem{Weitz1978}
D.A. Weitz, W.J. Skocpol, and M.~Tinkham.
\newblock {Characterization of niobium point contacts showing Josephson effects
  in the far infrared}.
\newblock {\em Journal of Applied Physics}, 49(9):4873, 1978.

\bibitem{Sochnikov2014}
I.~Sochnikov, L.~Maier, C.A. Watson, J.R. Kirtley, C.~Gould, G.~Tkachov, E.M.
  Hankiewicz, C.~Br\"{u}ne, H.~Buhmann, L.W. Molenkamp, and K.A. Moler.
\newblock {Nonsinusoidal Current-Phase Relationship in Josephson Junctions from
  the 3D Topological Insulator HgTe}.
\newblock {\em Physical Review Letters}, 114(6), 2015.

\bibitem{Tinkham2004}
M.~Tinkham.
\newblock {\em {Introduction to Superconductivity}}.
\newblock Dover Publications, 2004.

\bibitem{Russer1972}
P.~Russer.
\newblock {Influence of Microwave Radiation on Current-Voltage Characteristic
  of Superconducting Weak Links}.
\newblock {\em Journal of Applied Physics}, 43(4):2008, 1972.

\bibitem{Cleuziou2007}
J.-P. Cleuziou, W.~Wernsdorfer, S.~Andergassen, S.~Florens, V.~Bouchiat, Th.
  Ondar\c{c}uhu, and M.~Monthioux.
\newblock {Gate-Tuned High Frequency Response of Carbon Nanotube Josephson
  Junctions}.
\newblock {\em Physical Review Letters}, 99(11):117001, 2007.

\bibitem{Heersche2007}
H.B. Heersche, P.~Jarillo-Herrero, J.B. Oostinga, L.M.K. Vandersypen, and A.F.
  Morpurgo.
\newblock {Bipolar supercurrent in graphene.}
\newblock {\em Nature}, 446(7131):56--9, 2007.

\bibitem{Chauvin2006}
M.~Chauvin, P.~{Vom Stein}, H.~Pothier, P.~Joyez, M.~E. Huber, D.~Esteve, and
  C.~Urbina.
\newblock {Superconducting atomic contacts under microwave irradiation}.
\newblock {\em Physical Review Letters}, 97(6):1--4, 2006.

\bibitem{Cuevas2002}
J.C. Cuevas, J.~Heurich, A.~Mart\'{\i}n-Rodero, A.~{Levy Yeyati}, and
  G.~Sch\"{o}n.
\newblock {Subharmonic shapiro steps and assisted tunneling in superconducting
  point contacts.}
\newblock {\em Physical Review Letters}, 88(15):157001, 2002.

\bibitem{Dubos2001}
P.~Dubos, H.~Courtois, O.~Buisson, and B.~Pannetier.
\newblock {Coherent low-energy charge transport in a diffusive S-N-S junction.}
\newblock {\em Physical Review Letters}, 87(20):206801, 2001.

\bibitem{McCumber1968}
D.E. McCumber.
\newblock {Effect of ac Impedance on dc Voltage-Current Characteristics of
  Superconductor Weak-Link Junctions}.
\newblock {\em Journal of Applied Physics}, 39(7):3113, 1968.

\bibitem{Beenakker1991}
C.W.J. Beenakker and H.~van Houten.
\newblock {Josephson current through a superconducting quantum point contact
  shorter than the coherence length}.
\newblock {\em Physical Review Letters}, 66(23):3056--3059, 1991.

\bibitem{Note1}
This is similar to that reported for etched InSb nanowire devices \cite
  {Rokhinson2012}.

\bibitem{Houzet2013}
M.~Houzet, J.S. Meyer, D.M. Badiane, and L.I. Glazman.
\newblock {Dynamics of Majorana States in a Topological Josephson Junction}.
\newblock {\em Physical Review Letters}, 111(4):046401, 2013.

\bibitem{Alicea2012}
J.~Alicea.
\newblock {New directions in the pursuit of Majorana fermions in solid state
  systems.}
\newblock {\em Reports on progress in physics}, 75(7):076501, 2012.

\bibitem{Konig2007}
M.~K\"{o}nig, S.~Wiedmann, C.~Br\"{u}ne, A.~Roth, H.~Buhmann, L.W. Molenkamp,
  X.-L. Qi, and S.-C. Zhang.
\newblock {Quantum Spin Hall Insulator State in HgTe Quantum Wells}.
\newblock {\em Science}, 318(5851):766, 2007.

\bibitem{McMillan1968}
W.L. McMillan.
\newblock {Tunneling model of the superconducting proximity effect}.
\newblock {\em Physical Review}, 175(2):537--542, 1968.

\bibitem{Fagas2005}
G.~Fagas, G.~Tkachov, A.~Pfund, and K.~Richter.
\newblock {Geometrical enhancement of the proximity effect in quantum wires
  with extended superconducting tunnel contacts}.
\newblock {\em Physical Review B}, 71(22):224510, 2005.

\bibitem{Kopnin2014}
N.B. Kopnin, A.S. Mel'nikov, I.A. Sadovskyy, and V.M. Vinokur.
\newblock {Weak links in proximity-superconducting two-dimensional electron
  systems}.
\newblock {\em Physical Review B}, 89(8):081402, 2014.

\bibitem{Chauvin2005}
M.~Chauvin.
\newblock {\em {The Josephson Effect in Atomic Contacts}}.
\newblock PhD thesis, Univerist\'{e} Pierre et Marie Curie - Paris 6, 2005.

\bibitem{Heida1998}
J.P. Heida, B.J. van Wees, T.M. Klapwijk, and G.~Borghs.
\newblock {Nonlocal supercurrent in mesoscopic Josephson junctions}.
\newblock {\em Physical Review B}, 57(10):R5618--R5621, 1998.

\bibitem{Ledermann1999}
U.~Ledermann, A.~Fauch\`{e}re, and G.~Blatter.
\newblock {Nonlocality in mesoscopic Josephson junctions with strip geometry}.
\newblock {\em Physical Review B}, 59(14):R9027--R9030, 1999.

\bibitem{Note2}
As mentioned previously, the geometrical capacitance is small, and we neglect
  it here \cite {Oostinga2013}.

\bibitem{Averin1995}
D.~Averin and A.~Bardas.
\newblock {ac Josephson Effect in a Single Quantum Channel}.
\newblock {\em Physical Review Letters}, 75(9):1831--1834, 1995.

\bibitem{Goffman2000}
M.F. Goffman, R.~Cron, A.~{Levy Yeyati}, P.~Joyez, M.H. Devoret, D.~Esteve, and
  C.~Urbina.
\newblock {Supercurrent in atomic point contacts and Andreev states}.
\newblock {\em Physical Review Letters}, 85(1):170--173, 2000.

\bibitem{Chauvin2007}
M.~Chauvin, P.~vom Stein, D.~Esteve, C.~Urbina, J.~Cuevas, and A.~Yeyati.
\newblock {Crossover from Josephson to Multiple Andreev Reflection Currents in
  Atomic Contacts}.
\newblock {\em Physical Review Letters}, 99(6):067008, 2007.

\bibitem{Sau2012}
J.D. Sau, E.~Berg, and B.I. Halperin.
\newblock {On the possibility of the fractional ac Josephson effect in
  non-topological conventional superconductor-normal-superconductor junctions}.
\newblock 2012.

\end{thebibliography}

\end{document}